\documentstyle[preprint,aps]{revtex}

\begin{document}


\def \positronon{\bar\psi}
\def \electronon{\psi}

\def \Wnorm{W}
\def \Wms{{\cal{W}}}
\def \Wbar{{\overline{\Wms}}}

\def \PHI{{\mit\Phi}}

\def \pole{{\mit\Pi}}
\def \POLE#1#2{ {\pole_{#1}}^{#2} }

\def \teh{\pole\kern-0.525em\pole}

\def \amp{{\cal A}}
\def \amphat{\skew5\widehat{\amp}}
\def \cycamp{\skew5\widetilde{\amp}}

\def \Jhat{\skew5\widehat{J}}
\def \Jms{{\cal{J}}}
\def \Jmshat{{\widehat{\Jms}}}
\def \Jslash{\thinspace{\not{\negthinspace J}}}
\def \Jbar{\skew5\bar{\Jms}}

\def \Ims{{\cal{I}}}
\def \Ibar{\skew3\bar{\Ims}}
\def \Ihat{\skew5\widehat{I}}
\def \Xms{{\cal{X}}}
\def \Yms{{\cal{Y}}}

\def \psihat{\skew2\widehat{\psi}}
\def \psibarhat{\skew2\widehat{\skew2\bar{\psi}}}
\def \psibar{\skew3\bar{\psi}}
\def \Psibar{\bar{\Psi}}
\def \Psihat{\widehat{\Psi}}

\def \GAMMA{{\mit\Gamma}}
\def \GAMMAbar{\bar\GAMMA}

\def \Lam{\Lambda}


\def \dfour#1{  { {d^4{#1}}\over{(2\pi)^4} }\thinspace  }
\def \dn#1{  { {d^d{#1}}\over{(2\pi)^d} }\thinspace  }


\def \permsum#1#2{\sum_{{\cal{P}}(#1\ldots #2)}}
\def \cycsum#1#2{\sum_{{\cal{C}}(#1\ldots #2)}}
\def \braket#1#2{ \langle #1 \thinspace\thinspace #2 \rangle }
\def \tekarb#1#2{ {\braket{#1}{#2}}^{*} }
\def \bra#1{ \langle #1 | }
\def \ket#1{ | #1 \rangle }
\def \num{{\cal{N}}}
\def \Trace{\enspace{\rm Tr}\thinspace}

\def \eee{{e}}

\def \eps{\epsilon}
\def \vareps{\varepsilon}
\def \down{{}_\downarrow}
\def \up{{}_\uparrow}
\def \ts{\thinspace}
\newcommand {\beq}{\begin{equation}}
\newcommand {\eeq}{\end{equation}}
\newcommand {\beqa}{\begin{eqnarray}}
\newcommand {\eeqa}{\end{eqnarray}}
\newcommand {\LF}{\nonumber\\}

\def \zee{{\cal{Z}}}
\def \Poff{{\cal{P}}}
\def \Qoff{{\cal{Q}}}
\def \Koff{{\cal{K}}}
\def \half{ \hbox{$1\over2$} }

\def \link#1#2#3{ {{\braket{#1}{#3}}\over{\bra{#1}#2\ket{#3}}} }
\def \invlink#1#2#3{ {{\bra{#1}#2\ket{#3}}\over{\braket{#1}{#3}}} }

\def \linkstar#1#2#3{ { {{\braket{#1}{#3}}^{*}}
\over{{\bra{#1}#2\ket{#3}}^{*}}} }

\def \epsslash{\thinspace{\not{\negthinspace \eps}}}
\def \Poffslash{\thinspace{\not{\negthinspace\negthinspace\Poff}}}

\hyphenation{ap-pen-dix  spin-or  spin-ors}

\draft
\preprint{Fermilab--Pub--93/389-T}
\title{Multigluon Helicity Amplitudes Involving a Quark Loop}
\author{Gregory Mahlon \cite{byline}}
\address{Fermi National Accelerator Laboratory \\
P.O. Box 500 \\
Batavia, IL  60510 }
\date{December 1993}
\maketitle
\begin{abstract}
We apply the solution to the recursion relation for the
double-off-shell quark current to the
problem of computing one loop
amplitudes with an arbitrary number of gluons.  We are
able to compute amplitudes for photon-gluon scattering,
electron-positron annihilation to gluons, and gluon-gluon
scattering via a quark loop in the case of like-helicity
gluons.  In addition, we present the result for the
one-loop gluon-gluon scattering amplitude when one of the gluons
has opposite helicity from the others.
\end{abstract}


\section{Introduction}

In this paper we will extend the discussion of ``simple''
one-loop amplitudes involving an arbitrary number of
gauge bosons begun in Ref.~\cite{fourthpaper} to the
case of QCD.  In particular, we will evaluate the one-loop
corrections to the following processes:
\beq
\gamma g \rightarrow gg\ldots g
\label{photonglue}
\eeq
\beq
e^{+} e^{-} \rightarrow gg\ldots g
\label{makeglue}
\eeq
and
\beq
g g \rightarrow gg\ldots g,
\label{glueglue}
\eeq
for the case of like-helicity gluons.  In addition, we
will consider~(\ref{glueglue}) in the case where
one of the gluons has opposite helicity from the rest.
In this paper we will consider only those contributions
to~(\ref{glueglue}) that arise from diagrams containing
a closed quark loop.

In the indicated helicity
configurations, all of the above processes vanish at tree level.
Therefore, their one-loop corrections should be especially
simple.  In particular, these corrections must be
ultraviolet and infrared finite:  there is no
counterterm to absorb any ultraviolet divergence,
and no tree graphs with an additional soft gluons to
handle an infrared divergence.  Furthermore, since the
amplitude for
$q \bar q$ annihilation to $n$ like-helicity  gluons
vanishes at tree level~\cite{QQglueVANISH}, the Cutkosky rules
my be used to demonstrate that these particular one-loop
diagrams do not contain any cuts in the complex plane.

There are three major ideas which make this calculation
feasible.  First, we treat the gauge bosons in the theory
on an equal footing with the fermions by utilizing
the multispinor representation for the gauge
field [\ref{SCHWINGER}--\ref{Other}].
With a proper choice for the spinor basis, and a clever
selection of the gauge boson polarization spinors, great
simplifications are obtained in the expressions
involved in the computation of Feynman diagrams.
We utilize Weyl-van der Waerden spinors for this purpose.
A summary of our conventions and notations may be found
in Ref.~\cite{thirdpaper}.
Second, the many terms in a calculation are conveniently
organized into gauge-invariant sub-groups by the color
factorization property of QCD amplitudes~\cite{colorfactorization}.
Finally, we will use recursive
methods~[\ref{Third},\ref{BGrecursion},\ref{DYpapers}]
to obtain expressions for the sums of large groups of
Feynman diagrams.  In particular, we define currents which are
the sum of all tree graphs containing exactly
$n$ gauge bosons in some given helicity configuration.
Explicit closed-form solutions for these currents exist
for certain special helicity configurations.  Currents
with two off-shell particles will play an important role
in our discussion of loop amplitudes.

This paper is organized as follows.  In Sec.~II, we will
discuss the double-off-shell quark current.  This current
consists of a single quark line with both ends off shell
plus $n$ on-shell gluons.  All of the amplitudes discussed
in this paper are derived from this current by joining the
two ends of the quark line and performing the appropriate
integration.  Since we have obtained explicit solutions
for this current in the case of $n$ like-helicity gluons,
we are able to compute the corresponding
loop amplitudes in a straightforward manner.
We present our computation for some of these amplitudes
in Sec.~III.  The favorable form of the solution for the
double-off-shell quark current allows us to evaluate
the integrals exactly for arbitrary $n$.   We obtain
compact expressions for the amplitudes for
processes~(\ref{photonglue})--(\ref{glueglue}) in the
case of like-helicity gluons, and a somewhat more
complicated expression for the case~(\ref{glueglue})
containing a single
opposite-helicity gluon.  In the cases where these
results overlap those previously obtained by
Bern, Kosower, and Dixon~[\ref{String4and5},\ref{STRINGn}]
using string-based methods, we find agreement.
In particular, our computation proves
the conjecture made in
Ref.~\cite{String}, as well as producing additional new
amplitudes.
We conclude with a few closing remarks in section~IV.


\section{The double-off-shell quark currents}  \label{CURRENTS}

In this section we will present recursion relations
for currents consisting of a quark line plus $n$
gluons attached all possible ways.  In contrast to the
earlier quark currents presented by Berends and Giele \cite{BG},
both ends of the quark line will be off shell.  The generalizations
from the Berends and Giele relations are straightforward.
We are able to solve these recursion relations for the case
of an arbitrary number of like-helicity gluons.

We adopt the convention that, in any given diagram,
all of the momenta are
directed inward.  The gluons will
carry momenta $k_1, k_2, \ldots, k_n$ and color indices
$a_1, a_2, \ldots, a_n$.  We will denote the quark momentum by
$\Poff$ and its color by $i$.  The antiquark will have momentum $\Qoff$
and color $j$.
Let us denote the complete double-off-shell quark current by
$\Psihat_{ji}(\Qoff;1,\ldots,n;\Poff)$.
Momentum conservation implies that not all of the momenta
independent; indeed, we have
\beqa
\Poff &=& -[ \Qoff + k_1 + \cdots + k_n ]
\LF
&\equiv& -[\Qoff + \kappa(1,n)].
\eeqa
We will often suppress ``$\Poff$'' or ``$\Qoff$'' from the
argument list of $\Psihat$ when convenient.

The factorization of the
current into color and kinematical pieces
found by Berends and Giele \cite{BG} for the  quark current with
only one off-shell particle
still holds:
\beq
\Psihat_{ji}(\Qoff;1,\ldots,n;\Poff) =
g^n \permsum{1}{n}
( \Omega[1,n] )_{ji}
\Psi(\Qoff;1,\ldots,n;\Poff).
\label{colorfactorization}
\eeq
In Eq.~(\ref{colorfactorization}) we have used the notation
\beq
\Omega[1,n] \equiv
T^{a_1} T^{a_2} \cdots T^{a_n},
\eeq
where the $T^a$'s are color matrices in the fundamental representation
of the gauge group.  Because of our definitions, the complete
current $\Psihat$ is a
symmetric function of its gluon arguments, whereas
the order of the arguments appearing in $\Psi$ is important.
Hence, the current $\Psi$ is often referred to as the color-ordered
quark current.

The recursion relation satisfied by the color-ordered double-off-shell
quark current may be obtained in two different ways for each
possible chirality of the quark line.  The first way is to begin
with a current containing an off-shell antiquark and on-shell quark.
The quark may then be taken off mass shell to produce
the desired  double-off-shell current.  Derived in this manner,
we obtain the following recursion relation for the left-handed
quark current:
\beq
\Psi_{\alpha\dot\alpha}(\Qoff;1,\ldots,n;\Poff) =
-\sqrt2
\sum_{j=0}^{n-1}
\Psi_{\alpha\dot\beta}(\Qoff;1,\ldots,j)
\Jbar^{\dot\beta\beta}(j{+}1,\ldots,n)
{
{ [\Qoff + \kappa(1,n)]_{\beta\dot\alpha} }
\over
{ [\Qoff + \kappa(1,n)]^2 }
},
\label{LHQrecA}
\eeq
where $\Jms(1,\ldots,n)$
refers to the the color-ordered $n$-gluon
current (see Appendix).
The other option is to begin with a current containing an
off-shell quark and an on-shell antiquark.  This gives
\beq
\Psi_{\alpha\dot\alpha}(\Qoff;1,\ldots,n;\Poff) =
\sqrt2
\sum_{j=1}^{n}
{
{ [\kappa(1,n) + \Poff]_{\alpha\dot\beta} }
\over
{ [\kappa(1,n) + \Poff]^2 }
}
\Jbar^{\dot\beta\beta}(1,\ldots,j)
\Psi_{\beta\dot\alpha}(j{+}1,\ldots,n;\Poff)
\label{LHQrecB}
\eeq
The solutions to~(\ref{LHQrecA}) and~(\ref{LHQrecB}) are
equivalent, although
it may be easier to
use one form
instead of the other in certain situations.

The recursion relations for the right-handed quark
current are precisely what one would expect
given~(\ref{LHQrecA}) and~(\ref{LHQrecB}).  We may write
either
\beq
\Psibar^{\dot\alpha\alpha}(\Qoff;1,\ldots,n;\Poff) =
-\sqrt2
\sum_{j=0}^{n-1}
\Psibar^{\dot\alpha\beta}(\Qoff;1,\ldots,j)
\Jms_{\beta\dot\beta}(j{+}1,\ldots,n)
{
{ [\bar\Qoff + \bar\kappa(1,n)]^{\dot\beta\alpha} }
\over
{ [\Qoff + \kappa(1,n)]^2 }
}
\label{RHQrecA}
\eeq
or
\beq
\Psibar^{\dot\alpha\alpha}(\Qoff;1,\ldots,n;\Poff) =
\sqrt2
\sum_{j=1}^{n}
{
{ [ \bar\kappa(1,n) + \bar\Poff]^{\dot\alpha\beta} }
\over
{ [\kappa(1,n)+\Poff]^2 }
}
\Jms_{\beta\dot\beta}(1,\ldots,j)
\Psibar^{\dot\beta\alpha}(j{+}1,\ldots,n;\Poff).
\label{RHQrecB}
\eeq
The left- and right-handed  currents are connected by
the crossing relation
\beq
\Psibar^{\dot\alpha\alpha}(\Qoff;1,2,\ldots,n;\Poff) =
(-1)^{n+1}
\vareps^{\alpha\beta} \vareps^{\dot\alpha\dot\beta}
\Psi_{\beta\dot\beta}(\Poff;n,n{-}1,\ldots,1;\Qoff).
\label{QPSIcross}
\eeq

It is not difficult to solve the recursion relations~(\ref{LHQrecA})
and~(\ref{RHQrecA}) in the case of like-helicity gluons.  The gauge
choice suited to this case is the one given in
Ref.~\cite{BG}, namely
\beq
\eps_{\alpha\dot\alpha} =
{
{ u_{\alpha}(h) \bar u_{\dot\alpha}(k_j) }
\over
{ \braket{j}{h} }
}
\label{allpluspol}
\eeq
for the $j$th gluon.  The parameter $h$ appearing in (\ref{allpluspol})
is an arbitrary null vector which satisfies
\beq
h\cdot k_j \ne 0
\eeq
for all of the gluon momenta.
For the current, we find that
\beq
u^{\alpha}(h) \Psi_{\alpha\dot\alpha}(\Qoff;1^{+},\ldots,n^{+})
= u^{\alpha}(h) [\Qoff + \kappa(1,n)]_{\alpha\dot\alpha}
Y(\Qoff;1,\ldots,n)
\label{spinorizeLH}
\eeq
where the scalar function $Y$ is given by
\begin{mathletters}\label{Ysoln}
\beq
Y(\Qoff) =
{
{-i}
\over
{ \Qoff^2 }
}
\label{Ysolnzero}
\eeq
for zero gluons and
\beq
Y(\Qoff;1,\ldots,n) =
{
{ -i(-\sqrt2)^n }
\over
{ \bra{h} 1, \ldots, n \ket{h} }
}
\sum_{j=1}^n
u^{\beta}(h) \POLE{\beta}{\gamma}(\Qoff,1,\ldots,j) u_{\gamma}(h)
\label{Ysolnmult}
\eeq
for one or more gluons.
\end{mathletters}
The details of the  derivation
are similar to those given in Ref.~\cite{BG}.
The function $\pole$ appearing in~(\ref{Ysolnmult}) is given
by
\beq
\POLE{\beta}{\gamma}(\Qoff,1,\ldots,j) \equiv
{
{ k_{j\beta\dot\gamma}
  [\bar\Qoff + \bar\kappa(1,j)]^{\dot\gamma\gamma} }
\over
{ [\Qoff+\kappa(1,j{-}1)]^2 [\Qoff+\kappa(1,j)]^2 }
}.
\label{poledef}
\eeq
Note that is is possible to use momentum conservation
to eliminate $\Qoff$ from $Y$
in favor of $\Poff$, and then to use~(\ref{LHQrecB})
to obtain the complete expression for $\Psi_{\alpha\dot\alpha}$
[{\it{i.e.}}\ without contraction with $u^{\alpha}(h)$].
However, since that expression will not be required in the following,
we will not present it here.

Using the same gauge choice, the right-handed current is found to
be
\beq
\Psibar^{\dot\alpha\alpha}(\Qoff;1^{+},\ldots,n^{+})
u_{\alpha}(h)
= \bar\Qoff^{\dot\alpha\alpha} u_{\alpha}(h)
Y(\Qoff;1,\ldots,n),
\label{spinorizeRH}
\eeq
utilizing the same function $Y$.  It is a simple matter to
verify that the solutions~(\ref{spinorizeLH})
and~(\ref{spinorizeRH}) satisfy the crossing
relation~(\ref{QPSIcross}).

The forms given above are adequate if one wishes to study
tree-level processes.
Loop amplitudes, however, involve the integral of $Y$ (multiplied
by some other factors) over $\Qoff$.  Unless the other
factors contain inverse powers of $\Qoff$, we see that such
integrals will diverge.
It is easy to argue that the amplitudes for the processes
we are considering in this paper are ultraviolet finite
(in particular, because they vanish at tree-level).
Hence, any divergences are spurious.
To circumvent this difficulty, we introduce a ``regulated''
current, obtained by using the recursion relation to replace
$Y(\Qoff;1,\ldots,n)$ with an expression containing one
more propagator.
This new expression is then continued to $d$ dimensions and
simplified to the extent possible.
This  is the same method as the  one employed
in Ref.~\cite{fourthpaper} to solve the corresponding
problem in QED loop amplitudes, and the
steps required to derive the regulated current are
similar.  Hence, we will only present the result:
\beqa
Y(\Qoff;1,\ldots,n) &=&
{
{ -i(-\sqrt2)^n }
\over
{ \bra{h} 1, \ldots, n \ket{h} }
}
\Biggl\{
\sum_{j=1}^n
{
{ u^{\beta}(h)
  k_{j\beta\dot\gamma}
  [ \bar Q + \bar\kappa(1,j) ]^{\dot\gamma\gamma}
  u_{\gamma}(h) }
\over
{ \{[Q+\kappa(1,j{-}1)]^2-\mu^2 \}
  \{[Q+\kappa(1,j)]^2 -\mu^2 \} }
}
\LF &+&
\sum_{j=1}^{n-1}
{
{ \mu^2 \thinspace u^{\beta}(h)
  \kappa_{\beta\dot\gamma}(j{+}1,n)
  \bar k_j^{\dot\gamma\gamma}
  u_{\gamma}(h) }
\over
{ \{[Q+\kappa(1,j{-}1)]^2-\mu^2 \}
  \{[Q+\kappa(1,j)]^2 -\mu^2 \}
  \{[Q+\kappa(1,n)]^2 -\mu^2 \} }
}
\Biggr\}.
\LF &&
\label{regulatedY}
\eeqa
The new quantities $Q$ and $\mu^2$ appearing in~(\ref{regulatedY})
are a result of continuing the vector $\Qoff$ to
$d$ dimensions. They are defined as follows~\cite{Giele}:
$Q$ is a 4-dimensional vector consisting of the usual
space-time components of $\Qoff$, while
$\mu^2$ is the $(d-4)$-dimensional invariant
formed from the ``extra'' components of $\Qoff$,
\beq
-\mu^2 \equiv \Qoff^2 - Q^2.
\eeq
All of the other
momenta and polarization vectors remain in
4 dimensions, since they are external quantities~\cite{tHooftVeltman}.
The first term of~(\ref{regulatedY}) is what one would obtain
by simply regulating~(\ref{Ysolnmult}).  The presence of an additional
term reveals that
it is crucial to regulate {\it before}\ reducing the number
of propagators, to ensure that no illegal operations are performed
on divergent quantities.
For the processes considered in this paper, the integrals
obtained using~(\ref{regulatedY}) are convergent.  If this had not
been the case, then it would be necessary to apply the recursion
relation two (or more) times before regulating the expression.

\section{Construction of Loop Amplitudes}

In this section we will consider processes which may be studied
by joining the two ends of the double off-shell quark current
together at a vertex.


\subsection{The process $\gamma g \rightarrow gg\cdots g$}

The simplest amplitude which we may imagine
forming from the double-off-shell quark current is~(\ref{photonglue}).
Figure~\ref{photonfig} illustrates the contributions to
this process.  The photon is attached at the point where the
two ends of the quark current come together.  We may write
the integrand corresponding to this process as
\beq
\Ihat(\Qoff;1,\ldots,n{-}1,n_{\gamma}) =
\permsum{1}{n-1}
(-ie_q)
\gamma^{\xi}
\Psihat_{ji}(\Qoff;1,\ldots,n{-}1)
\eps_{\xi}(n).
\eeq
We write ``$n_{\gamma}$'' in the argument list of $\Ihat$ to
indicate that the photon has momentum $k_n$.  The
polarization vector for the photon is $\eps(n)$.
We take all external momenta to flow into the diagram,
$\kappa(1,n) = 0$.  Physical processes are obtained by
crossing.
The quark
circulating in the loop of this diagram has charge $e_q$.
Note that Fig.~\ref{photonfig} represents all of the one-loop
contributions to this process since photons do not couple directly
to gluons.

Inserting the color factorization~(\ref{colorfactorization})
for the quark current produces
\beq
\Ihat(\Qoff;1,\ldots,n{-}1,n_{\gamma}) =
-ie_q\thinspace g^{n-1}
\permsum{1}{n-1}
tr \thinspace \{ \Omega[1,n{-}1] \}
\thinspace\thinspace
{\rm Tr}\thinspace \{ \Psi(\Qoff;1,\ldots,n{-}1)
\epsslash\thinspace(n)\}.
\label{photonstart}
\eeq
Equation~(\ref{photonstart}) is valid for an arbitrary combination
of helicities.
The quark currents discussed in Sec.~\ref{CURRENTS} all have
like-helicity gluons, so that is the case we will discuss here.
To obtain the amplitude with a positive helicity photon and
$n-1$ positive helicity gluons, we make the gauge choice
given in~(\ref{allpluspol}), not only for the gluons, but for
the photon as well.  We will leave $h$ as a free parameter.
To obtain the amplitude with a negative helicity photon and
$n-1$ positive helicity gluons, we use~(\ref{allpluspol}) for
the gluons and set $h=k_n$.  For the photon, we write
\beq
\eps_{\alpha\dot\alpha}(n^{-}) =
{
{ u_{\alpha}(k_n) \bar u_{\dot\alpha}(h) }
\over
{ \tekarb{n}{h} }
\label{NegativeHelicity}
}.
\eeq
Because the computations are similar, we will concentrate
on the positive helicity photon case, and simply quote the
result for a negative helicity photon.

Before proceeding, however, let us pause to define a
color-ordered integrand and a color-ordered amplitude.
If we write
\beq
\Ihat(\Qoff;1^{+},\ldots,(n{-}1)^{+},n_{\gamma}^{+}) \equiv
-ie_q\thinspace g^{n-1}  \negthinspace \negthinspace \negthinspace
\permsum{1}{n-1}
tr \thinspace \{ \Omega[1,n{-}1] \}  \thinspace
I(\Qoff;1^{+},\ldots,(n{-}1)^{+},n_{\gamma}^{+})
\label{FullIntPhoton})
\eeq
then, obviously
\beq
I(\Qoff;1^{+},\ldots,(n{-}1)^{+},n_{\gamma}^{+}) =
{\rm Tr}\thinspace \{ \Psi(\Qoff;1^{+},\ldots,(n{-}1)^{+})
\epsslash\thinspace(n^{+})\}.
\label{PhotonStartOrdered}
\eeq
The actual amplitude is obtained from~(\ref{FullIntPhoton})
by integrating over $\Qoff$ and supplying the factor
$-1$ for closing the loop.  Likewise, we may obtain a color-ordered
amplitude from~(\ref{PhotonStartOrdered}) in analogous manner.

Let us convert the trace over Dirac matrices that appears
in~(\ref{photonstart}) to spinor notation, and substitute
for the photon polarization.
Since we are considering the massless limit, in which the two
possible chiralities for the circulating quark are decoupled,
we obtain two terms:
\beqa
I&&(\Qoff;1^{+},\ldots,(n{-}1)^{+},n_{\gamma}^{+}) =
\LF && \quad
{
{\sqrt2}
\over
{\braket{n}{h}}
}
\biggl\{
\bar u_{\dot\alpha}(k_n)
\Psibar^{\dot\alpha\alpha}(\Qoff;1^{+},\ldots,(n{-}1)^{+})
u_{\alpha}(h)
+ u^{\alpha}(h)
\Psi_{\alpha\dot\alpha}(\Qoff;1^{+},\ldots,(n{-}1)^{+})
\bar u^{\dot\alpha}(k_n)
\biggr\}.
\LF &&
\label{Spinored}
\eeqa
Notice that the undotted index of the quark current always
appears contracted with the gauge spinor, as promised in
Sec.~\ref{CURRENTS}.  Hence, we may use~(\ref{spinorizeLH})
and~(\ref{spinorizeRH}) for the spinor structure of the
currents:
\beq
I(\Qoff;1^{+},\ldots,(n{-}1)^{+},n_{\gamma}^{+}) =
2{\sqrt2} \thinspace
{
{ \bar u_{\dot\alpha}(k_n)
  \bar\Qoff^{\dot\alpha\alpha}
  u_{\alpha}(h) }
\over
{\braket{n}{h}}
}
Y(\Qoff;1,\ldots,n{-}1).
\label{NoMore}
\eeq
We have utilized the Weyl equation plus over-all
momentum conservation for the diagram to combine
the two contributions
from~(\ref{Spinored}) into a single term.

Since~(\ref{NoMore}) does not contain any extra
inverse powers of $\Qoff$, we must use the regulated
form of $Y$ given in~(\ref{regulatedY}), and
perform a $d$-dimensional integration.
The integrals which arise
during this procedure  have been
discussed in Ref.~\cite{fourthpaper} in connection
with $n$-photon scattering amplitude.  Hence, we will
proceed immediately to the result for the
color-ordered amplitude, which reads
\beqa
\amp&&(1^{+},\ldots,(n{-}1)^{+},n_{\gamma}^{+}) =
\LF && \quad
{
{(-\sqrt2)^n}
\over
{48\pi^2}
}
\sum_{\ell=1}^{n-2}
{
{ u^{\beta}(h)
  \kappa_{\beta\dot\beta}(\ell{+}1,n{-}1)
  \bar k^{\dot\beta\delta}_{\ell}
  u_{\delta}(h) }
\over
{ \bra{h}1,\ldots,n{-}1\ket{h}
  \braket{n}{h} }
}
\thinspace
\bar u_{\dot\alpha}(k_n)
[ \bar\kappa(1,\ell{-}1)-\bar\kappa(\ell{+}1,n)]^{\dot\alpha\alpha}
u_{\alpha}(h).
\eeqa
In order to prove the gauge invariance of the final result, we
will consider
\beq
\cycamp(1^{+},\ldots,(n{-}1)^{+},n_{\gamma}^{+}) \equiv
\cycsum{1}{n-1}
\amp(1^{+},\ldots,(n{-}1)^{+},n_{\gamma}^{+}) .
\eeq
Because the color factor appearing in the full amplitude is
invariant under cyclic permutations of $\{1,\ldots,n{-}1\}$,
we may replace the color-ordered amplitude
$\amp$ by $\cycamp / (n{-}1) $
without altering the value of the final result.
In general, to prove gauge invariance, we must consider all
of the terms which have the same color structure.
Thus, it will be  necessary to make the indicated replacement
to demonstrate that the expression we obtain is independent
of the gauge spinor.
The cyclic symmetry present in this case is
more restrictive than the permutation symmetry which
occurred in the
study of the $n$-photon scattering amplitude.
As a consequence, although a portion of the derivation
presented here is the same as Ref.~\cite{fourthpaper},
there are certain key differences.

We begin by multiplying the summand by
$\braket{n{-}1}{1} / \braket{n{-}1}{1} $
and applying the Schouten identity
to the combination
$u^{\beta}(h) \braket{n{-}1}{1}$.
We thus obtain
\beqa
&&\cycamp(1^{+},\ldots,(n{-}1)^{+},n_{\gamma}^{+}) =
\LF && \quad\enspace
{
{(-\sqrt2)^n}
\over
{48\pi^2}
}
\negthinspace\negthinspace\negthinspace
\cycsum{1}{n-1}
\sum_{\ell=1}^{n-2}
{
{ u^{\beta}(k_1)
  \kappa_{\beta\dot\beta}(\ell{+}1,n{-}1)
  \bar k^{\dot\beta\delta}_{\ell}
  u_{\delta}(h) }
\over
{ \bra{h}1,\ldots,n{-}1\ket{1}
  \braket{n}{h} }
}
\thinspace
\bar u_{\dot\alpha}(k_n)
[ \bar\kappa(1,\ell{-}1){-}\bar\kappa(\ell{+}1,n)]^{\dot\alpha\alpha}
u_{\alpha}(h)
\LF && \enspace +
{
{(-\sqrt2)^n}
\over
{48\pi^2}
}
\negthinspace\negthinspace\negthinspace
\cycsum{1}{n-1}
\sum_{\ell=1}^{n-2}
{
{ u^{\beta}(k_{n-1})
  \kappa_{\beta\dot\beta}(\ell{+}1,n{-}2)
  \bar k^{\dot\beta\delta}_{\ell}
  u_{\delta}(h) }
\over
{ \bra{n{-}1}1,\ldots,n{-}1\ket{h}
  \braket{n}{h} }
}
\thinspace
\bar u_{\dot\alpha}(k_n)
[ \bar\kappa(1,\ell{-}1){-}\bar\kappa(\ell{+}1,n)]^{\dot\alpha\alpha}
u_{\alpha}(h).
\LF &&
\label{CycleMe}
\eeqa
We take advantage of the sum over cyclic permutations by
relabeling the momenta in the second term of~(\ref{CycleMe})
as follows:
\beq
1 \rightarrow 2 \rightarrow 3 \rightarrow \cdots \rightarrow
n{-}1 \rightarrow 1.
\eeq
The effect of this relabeling is to produce
\beqa
\cycamp_2  =
{
{(-\sqrt2)^n}
\over
{48\pi^2}
}
\negthinspace\negthinspace\negthinspace
\cycsum{1}{n-1}
\sum_{\ell=1}^{n-2}  &&
{
{ u^{\beta}(k_1)
  \kappa_{\beta\dot\beta}(\ell{+}2,n{-}1)
  \bar k^{\dot\beta\delta}_{\ell+1}
  u_{\delta}(h) }
\over
{ \bra{1}2,\ldots,n{-}1,1\ket{h}
  \braket{n}{h} }
}
\LF && \times
\thinspace
\bar u_{\dot\alpha}(k_n)
[ \bar\kappa(2,\ell)-\bar\kappa(\ell{+}2,n{-}1)
  -\bar k_1 ]^{\dot\alpha\alpha}
u_{\alpha}(h).
\eeqa
A little bit of algebra converts this expression to
\beqa
\cycamp_2  = -
{
{(-\sqrt2)^n}
\over
{48\pi^2}
}
\negthinspace\negthinspace\negthinspace
\cycsum{1}{n-1}
\sum_{\ell=2}^{n-2}  &&
{
{ u^{\beta}(k_1)
  \kappa_{\beta\dot\beta}(\ell{+}1,n{-}1)
  \bar k^{\dot\beta\delta}_{\ell}
  u_{\delta}(h) }
\over
{ \bra{h}1,\ldots,n{-}1\ket{1}
  \braket{n}{h} }
}
\LF && \times
\thinspace
\bar u_{\dot\alpha}(k_n)
[ \bar\kappa(1,\ell{-}1)-\bar\kappa(\ell{+}1,n{-}1)
  -2\bar k_1 ]^{\dot\alpha\alpha}
u_{\alpha}(h).
\label{Cycled}
\eeqa
When~(\ref{Cycled}) is recombined with the first term
of~(\ref{CycleMe}), all but the $\ell=1$ term of that
sum cancels, and we are left with
\beqa
\cycamp(1^{+},\ldots,(n{-}1)^{+},n_{\gamma}^{+})  = &&
-{
{(-\sqrt2)^n}
\over
{48\pi^2}
}
\negthinspace\negthinspace\negthinspace
\cycsum{1}{n-1}
{
{ \bar u_{\dot\alpha}(k_n)
  \bar k^{\dot\alpha\beta}_{1}
  \kappa_{\beta\dot\beta}(2,n{-}1)
  \bar k^{\dot\beta\delta}_{1}
  u_{\delta}(h) }
\over
{ \bra{1}2,\ldots,n{-}1\ket{1}
  \braket{n}{h} }
}
\LF &&  -
{
{(-\sqrt2)^n}
\over
{48\pi^2}
}
\negthinspace\negthinspace\negthinspace
\cycsum{1}{n-1}
\sum_{\ell=1}^{n-2}
{
{ 2\thinspace\bar u_{\dot\alpha}(k_n)
  \bar k^{\dot\alpha\beta}_{1}
  \kappa_{\beta\dot\beta}(\ell{+}1,n{-}1)
  \bar k^{\dot\beta\delta}_{\ell}
  u_{\delta}(h) }
\over
{ \bra{1}2,\ldots,n{-}1\ket{1}
  \braket{n}{h} }
}.
\label{SuperCycleMe}
\eeqa
In the QED case, the first term of~(\ref{SuperCycleMe}) vanishes
when the sum over permutations of $\{2,\ldots,n{-}1\}$ is
performed.  In this case, however, we have but a cyclic
symmetry to work with, and this term must be retained.

Working within the allowed symmetry, we note
that the denominators appearing in~(\ref{SuperCycleMe}) are
invariant under cyclic permutations of $\{1,\ldots,n{-}1\}$.
We will exploit this symmetry to recast the sum  appearing
in the second term, by  relabeling the successive terms
in the sum on $\ell$ so that $k_{\ell}$ always becomes
$k_1$, $k_{\ell{+}1}$ always becomes $k_2$, etc.  In the
$\ell=2$ term, $k_1$ becomes $k_{n-1}$.  For $\ell=3$,
$k_1$ becomes $k_{n-2}$, and so on through $\ell=n-2$,
in which $k_1$ becomes $k_3$.  Consequently, we may
rewrite~(\ref{SuperCycleMe}) as
\beqa
\cycamp(1^{+},\ldots,(n{-}1)^{+},n_{\gamma}^{+})  = &&
-{
{(-\sqrt2)^n}
\over
{48\pi^2}
}
\negthinspace\negthinspace\negthinspace
\cycsum{1}{n-1}
{
{ \bar u_{\dot\alpha}(k_n)
  \bar k^{\dot\alpha\beta}_{1}
  \kappa_{\beta\dot\beta}(2,n{-}1)
  \bar k^{\dot\beta\delta}_{1}
  u_{\delta}(h) }
\over
{ \bra{1}2,\ldots,n{-}1\ket{1}
  \braket{n}{h} }
}
\LF &&  -
{
{(-\sqrt2)^n}
\over
{48\pi^2}
}
\negthinspace\negthinspace\negthinspace
\cycsum{1}{n-1}
\sum_{j=3}^{n-1}
{
{ 2\thinspace\bar u_{\dot\alpha}(k_n)
  \bar k^{\dot\alpha\beta}_{j}
  \kappa_{\beta\dot\beta}(2,j{-}1)
  \bar k^{\dot\beta\delta}_{1}
  u_{\delta}(h) }
\over
{ \bra{1}2,\ldots,n{-}1\ket{1}
  \braket{n}{h} }
}.
\label{SuperCycled}
\eeqa

The next step is to use the anticommutation properties of the
$\sigma$-matrices to write
\beqa
{ 2\thinspace\bar u_{\dot\alpha}(k_n)
  \bar k^{\dot\alpha\beta}_{j}
  \kappa_{\beta\dot\beta}(2,j{-}1)
  \bar k^{\dot\beta\delta}_{1}
  u_{\delta}(h) }  &=&
{ \bar u_{\dot\alpha}(k_n)
  \bar k^{\dot\alpha\beta}_{j}
  \kappa_{\beta\dot\beta}(2,j{-}1)
  \bar k^{\dot\beta\delta}_{1}
  u_{\delta}(h) }
\LF &-&
{ \bar u_{\dot\alpha}(k_n)
  \bar\kappa^{\dot\alpha\beta}(2,j{-}1)
  k_{j\beta\dot\beta}
  \bar k^{\dot\beta\delta}_{1}
  u_{\delta}(h) }
\LF &+&
{ 2\thinspace k_j \cdot \kappa(2,j{-}1) \thinspace\thinspace
  \bar u_{\dot\beta}(k_n)
  \bar k^{\dot\beta\delta}_{1}
  u_{\delta}(h) } .
\label{Anticommuted}
\eeqa
It is a simple exercise to demonstrate that the contribution generated
from the last term of~(\ref{Anticommuted}) exactly cancels
the first term of~(\ref{SuperCycled}).  We will designate the
contributions from the first and second terms~(\ref{Anticommuted})
$\cycamp_1$ and $\cycamp_2$ respectively.

We will consider $\cycamp_2$ first, since we will perform the
fewest operations on it.  From~(\ref{Anticommuted})
and~(\ref{SuperCycled}) we see that
\beqa
\cycamp_2  = &&
{
{(-\sqrt2)^n}
\over
{48\pi^2}
}
\negthinspace\negthinspace\negthinspace
\cycsum{1}{n-1}
\sum_{j=3}^{n-1}
{
{ \bar u_{\dot\alpha}(k_n)
  \bar\kappa^{\dot\alpha\beta}(2,j{-}1)
  k_{j\beta\dot\beta}
  \bar k^{\dot\beta\delta}_{1}
  u_{\delta}(h) }
\over
{ \bra{1}2,\ldots,n{-}1\ket{1}
  \braket{n}{h} }
}.
\label{A2start}
\eeqa
If we supply the factor $\braket{1}{n} / \braket{1}{n}$
and perform a bit of spinor algebra, we find
\beqa
\cycamp_2  = &&
{
{(-\sqrt2)^n}
\over
{48\pi^2}
}
\negthinspace\negthinspace\negthinspace
\cycsum{1}{n-1}
\sum_{j=3}^{n-1}
{
{ \bar k_j^{\dot\beta\beta}
  \kappa_{\beta\dot\alpha}(2,j{-}1)
  \bar k_{n}^{\dot\alpha\delta}
  k_{1\delta\dot\beta} }
\over
{ \bra{1}2,\ldots,n{-}1\ket{1} }
}
\link{1}{n}{h}.
\label{A2}
\eeqa
We will save~(\ref{A2}) for later cancellation.

Turning to $\cycamp_1$ and supplying a factor
$\braket{1}{n} / \braket{1}{n}$ as we did for the other
term, we find that it may be written
\beq
\cycamp_1  =
-{
{(-\sqrt2)^n}
\over
{48\pi^2}
}
\negthinspace\negthinspace\negthinspace
\cycsum{1}{n-1}
\sum_{j=3}^{n-1}
{
{  \bar u_{\dot\beta}(k_1)
   \bar\kappa^{\dot\beta\beta}(1,j{-}1)
   k_{j\beta\dot\alpha}
   \bar k_n^{\dot\alpha\alpha}
   u_{\alpha}(k_1) }
\over
{ \bra{1}2,\ldots,n{-}1\ket{1} }
}
\link{1}{n}{h}.
\eeq
We now break $\cycamp_1$ into two pieces, utilizing momentum
conservation and the Weyl equation to write
\beqa
k_{j\beta\dot\alpha}
\bar k_n^{\dot\alpha\alpha}
u_{\alpha}(k_1)
&=& -
k_{j\beta\dot\alpha}
\bar \kappa^{\dot\alpha\alpha}(2,n{-}1)
u_{\alpha}(k_1)
\LF
&=& -
k_{j\beta\dot\alpha}
\bar \kappa^{\dot\alpha\alpha}(2,j{-}1)
u_{\alpha}(k_1)
- k_{j\beta\dot\alpha}
\bar \kappa^{\dot\alpha\alpha}(j{+}1,n{-}1)
u_{\alpha}(k_1).
\label{CleverSplit}
\eeqa
The second contribution in the last line of~(\ref{CleverSplit})
vanishes if $j = n-1$.   Thus, we may write
\beqa
\cycamp_1 &&  =
{
{(-\sqrt2)^n}
\over
{48\pi^2}
}
\negthinspace\negthinspace\negthinspace
\cycsum{1}{n-1}
\sum_{j=3}^{n-1}
\sum_{\ell=2}^{j-1}
{
{  \bar u_{\dot\beta}(k_1)
   \bar\kappa^{\dot\beta\beta}(1,j{-}1)
   k_{j\beta\dot\alpha}
   \bar k_{\ell}^{\dot\alpha\alpha}
   u_{\alpha}(k_1) }
\over
{ \bra{1}2,\ldots,n{-}1\ket{1} }
}
\link{1}{n}{h}
\LF &&
+{
{(-\sqrt2)^n}
\over
{48\pi^2}
}
\negthinspace\negthinspace\negthinspace
\cycsum{1}{n-1}
\sum_{j=2}^{n-2}
\sum_{\ell=j+1}^{n-1}
{
{  \bar u_{\dot\beta}(k_1)
   \bar\kappa^{\dot\beta\beta}(1,j{-}1)
   k_{j\beta\dot\alpha}
   \bar k_{\ell}^{\dot\alpha\alpha}
   u_{\alpha}(k_1) }
\over
{ \bra{1}2,\ldots,n{-}1\ket{1} }
}
\link{1}{n}{h},
\label{A1cont}
\eeqa
where the extension of the second sum to include $j=2$ may be
done without producing a compensating term since the contribution
in question vanishes.  We have also chosen to write the
implicit $\kappa$-sums from~(\ref{CleverSplit}) explicitly
as sums over $\ell$.

The next step is to interchange the order of the double sum
appearing in the first term of~(\ref{A1cont}).   If we
subsequently interchange the names of the
two dummy summation variables, we see that the summation
ranges of the two terms are identical.  Performing this
manipulation and rearranging the spinor products a bit we
find
\beqa
\cycamp_1 && =
{
{(-\sqrt2)^n}
\over
{48\pi^2}
}
\negthinspace\negthinspace\negthinspace
\cycsum{1}{n-1}
\sum_{j=2}^{n-2}
\sum_{\ell=x+1}^{n-1}
{
{ \bar k_j^{\dot\alpha\alpha}
  k_{1\alpha\dot\beta}
  \bar\kappa^{\dot\beta\beta}(1,\ell{-}1)
  k_{\ell\beta\dot\alpha} }
\over
{ \bra{1}2,\ldots,n{-}1\ket{1} }
}
\link{1}{n}{h}
\LF &&
+{
{(-\sqrt2)^n}
\over
{48\pi^2}
}
\negthinspace\negthinspace\negthinspace
\cycsum{1}{n-1}
\sum_{j=2}^{n-2}
\sum_{\ell=j+1}^{n-1}
{
{ \bar k_j^{\dot\alpha\alpha}
  \kappa_{\alpha\dot\beta}(2,j{-}1)
  \bar k^{\dot\beta\beta}_{1}
  k_{\ell\beta\dot\alpha} }
\over
{ \bra{1}2,\ldots,n{-}1\ket{1} }
}
\link{1}{n}{h}.
\label{splut}
\eeqa
To make the terms of~(\ref{splut})
look even more nearly alike, we write
\beq
k_{1\alpha\dot\beta} = \kappa_{\alpha\dot\beta}(1,j{-}1)
                     - \kappa_{\alpha\dot\beta}(2,j{-}1)
\eeq
in the first term
and
\beq
\bar k_1^{\dot\beta\beta}
= \bar\kappa^{\dot\beta\beta}(1,\ell{-}1)
- \bar\kappa^{\dot\beta\beta}(2,\ell{-}1)
\eeq
in the second term.  Two of the four terms thus generated  cancel,
leaving only
\beqa
\cycamp_1 && =
{
{(-\sqrt2)^n}
\over
{48\pi^2}
}
\negthinspace\negthinspace\negthinspace
\cycsum{1}{n-1}
\sum_{j=2}^{n-2}
\sum_{\ell=x+1}^{n-1}
{
{ \bar k_j^{\dot\alpha\alpha}
  \kappa_{\alpha\dot\beta}(1,j{-}1)
  \bar\kappa^{\dot\beta\beta}(1,\ell{-}1)
  k_{\ell\beta\dot\alpha} }
\over
{ \bra{1}2,\ldots,n{-}1\ket{1} }
}
\link{1}{n}{h}
\LF &&
-{
{(-\sqrt2)^n}
\over
{48\pi^2}
}
\negthinspace\negthinspace\negthinspace
\cycsum{1}{n-1}
\sum_{j=2}^{n-2}
\sum_{\ell=j+1}^{n-1}
{
{ \bar k_j^{\dot\alpha\alpha}
  \kappa_{\alpha\dot\beta}(2,j{-}1)
  \bar \kappa^{\dot\beta\beta}(2,\ell{-}1)
  k_{\ell\beta\dot\alpha} }
\over
{ \bra{1}2,\ldots,n{-}1\ket{1} }
}
\link{1}{n}{h}.
\label{splutted}
\eeqa

We will designate the first term of~(\ref{splutted})
$\cycamp_{1{\rm A}}$ and save it in its present form.
The other contribution will be referred to as
$\cycamp_{1{\rm B}}$.   We  begin operations on it
by using the cyclic sum to shift all of the dummy indices
down by one unit, with $k_1 \rightarrow k_{n-1}$.
The result reads
\beq
\cycamp_{1{\rm B}}  =
-{
{(-\sqrt2)^n}
\over
{48\pi^2}
}
\negthinspace\negthinspace\negthinspace
\cycsum{1}{n-1}
\sum_{j=2}^{n-2}
\sum_{\ell=j+1}^{n-1}
{
{ \bar k_{j-1}^{\dot\alpha\alpha}
  \kappa_{\alpha\dot\beta}(1,j{-}2)
  \bar \kappa^{\dot\beta\beta}(1,\ell{-}2)
  (k_{\ell-1})_{\beta\dot\alpha} }
\over
{ \bra{1}2,\ldots,n{-}1\ket{1} }
}
\link{n{-}1}{n}{h}.
\label{A1Bstart}
\eeq
Shifting the summations by 1 unit
produces
\beq
\cycamp_{1{\rm B}}  =
-{
{(-\sqrt2)^n}
\over
{48\pi^2}
}
\negthinspace\negthinspace\negthinspace
\cycsum{1}{n-1}
\sum_{j=2}^{n-3}
\sum_{\ell=j+1}^{n-2}
{
{ \bar k_{j}^{\dot\alpha\alpha}
  \kappa_{\alpha\dot\beta}(1,j{-}1)
  \bar \kappa^{\dot\beta\beta}(1,\ell{-}1)
  k_{\ell\beta\dot\alpha} }
\over
{ \bra{1}2,\ldots,n{-}1\ket{1} }
}
\link{n{-}1}{n}{h}.
\label{A1B}
\eeq
Since
\beq
\link{1}{n}{h} - \link{n{-}1}{n}{h} = -\link{n{-}1}{n}{1},
\label{FinalCombo}
\eeq
it is desirable to extend the summation range appearing
in~(\ref{A1B}) to match the summation range of the first
term in~(\ref{splutted}).  This time, the compensating
contribution does not vanish, but instead reads
\beq
\cycamp_{1{\rm X}}  \equiv
{
{(-\sqrt2)^n}
\over
{48\pi^2}
}
\negthinspace\negthinspace\negthinspace
\cycsum{1}{n-1}
\sum_{j=2}^{n-2}
{
{ \bar k_{j}^{\dot\alpha\alpha}
  \kappa_{\alpha\dot\beta}(1,j{-}1)
  \bar \kappa^{\dot\beta\beta}(1,n{-}2)
  (k_{n-1})_{\beta\dot\alpha} }
\over
{ \bra{1}2,\ldots,n{-}1\ket{1} }
}
\link{n{-}1}{n}{h}.
\label{A1comp}
\eeq
If we apply the cyclic sum to~(\ref{A1comp}),
shifting the dummy momentum labels up by 1 unit, and then
shift the sum over $j$, we find that
\beq
\cycamp_{1{\rm X}}  =
{
{(-\sqrt2)^n}
\over
{48\pi^2}
}
\negthinspace\negthinspace\negthinspace
\cycsum{1}{n-1}
\sum_{j=3}^{n-1}
{
{ \bar k_{j}^{\dot\alpha\alpha}
  \kappa_{\alpha\dot\beta}(2,j{-}1)
  \bar \kappa^{\dot\beta\beta}(2,n{-}1)
  k_{1\beta\dot\alpha} }
\over
{ \bra{1}2,\ldots,n{-}1\ket{1} }
}
\link{1}{n}{h}.
\eeq
A straightforward application of momentum conservation plus
the Weyl equation reveals that $\cycamp_{1{\rm X}}$ is exactly
the term required to cancel $\cycamp_2$ [see Eq.~(\ref{A2})].
Thus, the entire result reads
\beq
\cycamp(1^{+},\ldots,(n{-}1)^{+},n_{\gamma}^{+})  =
-{
{(-\sqrt2)^n}
\over
{48\pi^2}
}
\negthinspace\negthinspace\negthinspace
\cycsum{1}{n-1}
\sum_{j=2}^{n-2}
\sum_{\ell=j+1}^{n-1}
{
{ \bar k_{j}^{\dot\alpha\alpha}
  \kappa_{\alpha\dot\beta}(1,j)
  \bar \kappa^{\dot\beta\beta}(1,\ell)
  k_{\ell\beta\dot\alpha} }
\over
{ \bra{1}2,\ldots,n\ket{1} }
},
\eeq
implying the  amplitude
\beqa
\amphat&&(1^{+},\ldots,(n{-}1)^{+},n_{\gamma}^{+})  =
\LF && \quad
{
{i}
\over
{48\pi^2}
}
(-e_q \sqrt2)(-g\sqrt2)^{n-1}
\negthinspace\negthinspace\negthinspace\negthinspace
\permsum{1}{n-1} \negthinspace
tr \thinspace \{ \Omega[1,n{-}1] \}  \thinspace
\sum_{j=2}^{n-2}
\sum_{\ell=j+1}^{n-1}
{
{ \bar k_{j}^{\dot\alpha\alpha}
  \kappa_{\alpha\dot\beta}(1,j)
  \bar \kappa^{\dot\beta\beta}(1,\ell)
  k_{\ell\beta\dot\alpha} }
\over
{ \bra{1}2,\ldots,n\ket{1} }
}.
\LF &&
\label{PhotonGluon+}
\eeqa

It is easy to repeat the above calculation for a negative
helicity photon and $n-1$ positive helicity gluons.
We obtain
\beqa
\amphat&&(1^{+},\ldots,(n{-}1)^{+},n_{\gamma}^{-})  =
\LF && \qquad
{
{i}
\over
{48\pi^2}
}
(-e_q \sqrt2)(-g\sqrt2)^{n-1}
\negthinspace\negthinspace\negthinspace\negthinspace
\permsum{1}{n-1} \negthinspace
tr \thinspace \{ \Omega[1,n{-}1] \}  \thinspace
\LF && \qquad
\qquad\qquad\qquad\qquad\qquad\qquad
\times
\sum_{j=2}^{n-2}
\sum_{\ell=j+1}^{n-1}
{
{ k_{j\alpha\dot\alpha}
  \bar\kappa^{\dot\alpha\beta}(1,j)
  \kappa_{\beta\dot\beta}(1,\ell)
  \bar k_{\ell}^{\dot\beta\alpha} }
\over
{ \bra{1}2,\ldots,n{-}1\ket{1} }
}
\linkstar{1}{n}{n{-}1}.
\LF &&
\label{PhotonGluon-}
\eeqa
This calculation proceeds in almost exactly the same manner
as the positive helicity photon case, except that the
step corresponding to Eq.~(\ref{SuperCycled}) should be skipped.

The amplitudes presented in Eqs.~(\ref{PhotonGluon+})
and~(\ref{PhotonGluon-}) should reduce to the amplitudes
for $n$-photon scattering in the $U(1)$ limit.  This
limit is simple to obtain:  simply replace all of the color
matrices by the unit matrix.  We have verified
that the above expressions do indeed reduce to the values for
the $n$-photon amplitude reported in Ref.~\cite{fourthpaper}.


\subsection{The process $e^{+} e^{-} \rightarrow gg\cdots g$}

The lowest order diagrams for electron-positron annihilation
to gluons are
obtained from Fig.~\ref{photonfig} by attaching an
electron-positron pair to the photon.  There are no
additional diagrams
because  leptons
do not couple directly to gluons.
This process is closely related to the one examined in the
last section:  in the
limit where the  $e^{+}e^{-}$ pair becomes collinear it is
simply some (divergent) factor times the amplitude obtained above.
Thus, it is not surprising that the entire discussion of
the last section may be applied to this case with very little
modification.

For a negative helicity positron of momentum $p$, a positive
helicity electron of momentum $q$, and $n$ positive helicity
gluons we find
\beqa
\amphat&&(p^{-};1^{+},\ldots,(n)^{+};q^{+})  =
\LF && \quad
-{
{iee_q}
\over
{24\pi^2}
}
(-g\sqrt2)^{n}
\negthinspace\negthinspace\negthinspace\negthinspace
\permsum{1}{n} \negthinspace
tr \thinspace \{ \Omega[1,n] \}  \thinspace
\sum_{j=2}^{n-1}
\sum_{\ell=j+1}^{n}
{
{ k_{j\alpha\dot\alpha}
  \bar\kappa^{\dot\alpha\beta}(1,j)
  \kappa_{\beta\dot\beta}(1,\ell)
  \bar k_{\ell}^{\dot\beta\alpha} }
\over
{ \bra{1}2,\ldots,n\ket{1} \braket{p}{q} }
}
\linkstar{1}{q}{n}
\label{e+e-Gluons}
\eeqa
(all momenta directed inward).
To obtain the amplitude for a positive helicity positron of
momentum $p$, a negative helicity electron of momentum $q$,
and $n$ positive helicity gluons, interchange $p$ and $q$
in the above expression.


\subsection{The process $gg \rightarrow gg\cdots g$}

We now turn to the process of gluon-gluon scattering.  At
present, we have only been able to evaluate those diagrams
which contain a quark loop.
In principle, the diagrams involving a gluon loop should
be obtainable from a gluon current with two off-shell
gluons.  However, the complicated form of the recursion
relation for this object~\cite{thirdpaper}
makes this path difficult to follow.
Nevertheless, it is possible to
use  supersymmetry~\cite{SUSYrelations}  to
obtain some of the subamplitudes where a gluon
loop replaces the quark loop.

Fig.~\ref{gluegluefig}  illustrates the process in terms of
the double off-shell quark current and the gluon current
(single off-shell particle).  To avoid over-counting
we will ``anchor'' the $n$th gluon and write the integrand
as
\beq
\Ihat(\Qoff;1,\ldots,n) =
\permsum{1}{n-1} \sum_{t=1}^{n-1}
{
{1}
\over
{t! (n{-}t{-}1)! }
}
(-ig)(T^x)_{ij}
\gamma^{\xi}
\Psihat_{ji}(\Qoff;1,\ldots,t)
\Jhat^{x}_{\xi}(t{+}1,\ldots,n).
\label{gluonstart}
\eeq
In specifying that the sum on $t$ begin at $1$ instead of $0$,
we have dropped the (vanishing) tadpole diagram.
Insertion of the color factorizations for the quark and gluon
currents [Eqs.~(\ref{colorfactorization})
and~(\ref{factorizeglue})] produces
\beqa
\Ihat(\Qoff;1,\ldots,n) =
-2ig^n  \negthinspace \negthinspace \negthinspace
\permsum{1}{n-1} \sum_{t=1}^{n-1} \sum_{s=t}^{n-1} &&
tr \thinspace \{ \Omega[1,t]T^x \}
\thinspace\thinspace tr\{T^x \Omega[t{+}1,s] T^{a_n}
\Omega[s{+}1,n{-}1]\}
\LF && \negthinspace\times
{\rm Tr}\thinspace \{ \Psi(\Qoff;1,\ldots,t)
\Jslash(t{+}1,\ldots,s,n,s{+}1,\ldots,n{-}1)\}.
\LF &&
\label{integrandstart}
\eeqa

Let us examine the color factor appearing in~(\ref{integrandstart}):
\beq
C \equiv
tr \thinspace \{ \Omega[1,t]T^x \}
\thinspace\thinspace tr\{T^x \Omega[t{+}1,s] T^{a_n}
\Omega[s{+}1,n{-}1]\}
\eeq
We simplify it
by applying the completeness relation for
$SU(N)$ to do the implied sum on $x$.  The result,
\beq
C =
{ {1}\over{2} }
\thinspace tr \thinspace
\{ \Omega[s{+}1,n{-}1] \Omega[1,s] T^{a_n} \}
- { {1}\over{2N} }
\thinspace tr \thinspace
\{ \Omega[1,t] \}
\thinspace\thinspace tr \thinspace
\{ \Omega[s{+}1,n{-}1] \Omega[t{+}1,s] T^{a_n} \}
\label{ColorFactor}
\eeq
suggests that the manner in which we chose to label the
gluons when writing down~(\ref{gluonstart})  ({\it i.e.}\
the gluons labeled 1 through $t$
as a part of the quark current, and the
remaining  gluons labeled $t{+}1$ through $s$, $n$,
and then $s{+}1$ through $n{-}1$
as part of the gluon current) is not
the best way to proceed.  Consider the contribution to the
integrand from the second term of~(\ref{ColorFactor}).
The structure of this term suggests that we write
\beqa
\Ihat_2(\Qoff;1,\ldots,n) =
{{ig^n}\over{N}}  \negthinspace
\permsum{1}{n-1} \sum_{t=1}^{n-1}  \sum_{v=t}^{n-1} &&
tr \thinspace \{ \Omega[1,t] \}
\thinspace\thinspace tr\{ \Omega[t{+}1,n] \}
\LF && \negthinspace\times
{\rm Tr}\thinspace \{ \Psi(\Qoff;1,\ldots,t)
\Jslash(v{+}1,\ldots,n{-}1,n,t{+}1,\ldots,v)\},
\LF &&
\label{suppressed}
\eeqa
making use of the sum over permutations to simplify the
color factor.
In this form, it is obvious that we may apply the cyclic
sum identity \cite{BG}
\beq
\cycsum{t{+}1}{n} J_{\xi}(t{+}1,\ldots,n) = 0
\eeq
to show that all except the $t=n{-}1$ contribution
vanishes.  However, the color factor for this piece is
\beq
tr \thinspace \{ \Omega[1,n{-}1] \}
\thinspace\thinspace tr\{ T^{a_n} \}  = 0.
\eeq
Hence, the color-suppressed contribution to the amplitude vanishes.
Alternatively, we could have considered an extended $U(N)$
gauge theory.
The observation
that the gauge boson coupling the quark loop to the gluon
current can not have $U(1)$ quantum numbers in the extended
theory guarantees that the $U(N)$ and $SU(N)$ results coincide.
Then,  the absence of a $1/N$ term in the $U(N)$ completeness relation
implies the absence of such a term in the amplitude.

Turning to the remaining contribution from~(\ref{ColorFactor}),
we see that the order of the color
matrices suggests that we write
\beqa
\Ihat(\Qoff;1,\ldots,n) =
-ig^n  \negthinspace \negthinspace \negthinspace
\permsum{1}{n-1} &&
tr \thinspace \{ \Omega[1,n] \}
\LF && \times
\sum_{y=1}^{n-1} \sum_{z=0}^{y-1}
{\rm Tr}\thinspace \{ \Psi(\Qoff;z{+}1,\ldots,y)
\Jslash(y{+}1,\ldots,n,1,\ldots,z)\}.
\label{GenStart}
\eeqa
Equation~(\ref{GenStart}) may be taken as a general starting
point for {\it all}\ gluon-gluon scattering diagrams containing
a single quark loop.  It is valid for any helicity combination.
All that is required to utilize it is a knowledge of the
color-ordered currents appearing on the right hand side.

At this stage we define the color-ordered
integrand
and color-ordered amplitude.
If we write
\beq
\Ihat(\Qoff;1,\ldots,n) \equiv
-ig^n  \negthinspace \negthinspace \negthinspace
\permsum{1}{n-1}
tr \thinspace \{ \Omega[1,n] \}  \thinspace
I(\Qoff;1,\ldots,n)
\label{IntFactorize}
\eeq
then, obviously
\beq
I(\Qoff;1,\ldots,n) =
\sum_{y=1}^{n-1} \sum_{z=0}^{y-1}
{\rm Tr}\thinspace \{ \Psi(\Qoff;z{+}1,\ldots,y)
\Jslash(y{+}1,\ldots,n,1,\ldots,z)\}.
\label{IntOrdered}
\eeq
As before,
the complete amplitude
is obtained from~(\ref{IntFactorize}) by integrating over
$\Qoff$ and supplying the factor $-1$ for closing the loop.
The color-ordered amplitude is obtained
from~(\ref{IntOrdered}) in the same manner.

We now specialize to the
first
case for which we have obtained explicit
results:  $n$ like-helicity gluons.
In spinor notation,~(\ref{IntOrdered}) becomes
\beqa
I(\Qoff;1,\ldots,n) = {\sqrt2}
\sum_{y=1}^{n-1} \sum_{z=0}^{y-1}
\bigg\{  &&
\Psibar^{\dot\alpha\alpha}(\Qoff;z{+}1,\ldots,y)
\Jms_{\alpha\dot\alpha}(y{+}1,\ldots,n,1,\ldots,z)
\LF && +
\Psi_{\alpha\dot\alpha}(\Qoff;z{+}1,\ldots,y)
\Jbar^{\dot\alpha\alpha}(y{+}1,\ldots,n,1,\ldots,z)
\biggr\}.
\label{GOspinor}
\eeqa
We choose the gauge
indicated by~(\ref{allpluspol}).  Inserting the
solution~(\ref{Jallplus}) for
the gluon current  we find
\beqa
I(\Qoff;1,\ldots,n) =
\sum_{y=1}^{n-1} \sum_{z=0}^{y-1}  &&
{
{ -(-\sqrt2)^{n-y+z} }
\over
{ \bra{h} y{+}1,\ldots,n,1,2,\ldots,z \ket{h} }
}
\LF &&  \times
\bigg\{
u^{\beta}(h)[ \kappa(y{+}1,n)+\kappa(1,z) ]_{\beta\dot\alpha}
\Psibar^{\dot\alpha\alpha}(\Qoff;z{+}1,\ldots,y)
u_{\alpha}(h)
\LF && \quad -
u^{\alpha}(h)
\Psi_{\alpha\dot\alpha}(\Qoff;z{+}1,\ldots,y)
[ \bar\kappa(y{+}1,n)+\bar\kappa(1,z) ]^{\dot\alpha\beta}
u_{\beta}(h)
\biggr\}.
\label{Jsubbed}
\eeqa
Once again, as promised in Sec.~\ref{CURRENTS}, all occurrences
of the double-off-shell quark current have its
undotted index contracted into the
gauge spinor.  Using Eqs.~(\ref{spinorizeLH})
and~(\ref{spinorizeRH}) to insert the spinor structure
of the quark currents, plus over-all momentum conservation
for the diagram, we see that the two contributions to~(\ref{Jsubbed})
are actually equal.  That is,
\beq
I(\Qoff;1,\ldots,n) =
\sum_{y=1}^{n-1} \sum_{z=0}^{y-1}
{ (-\sqrt2)^{n-y+z} }
{
{ 2u^{\beta}(h)\kappa_{\beta\dot\alpha}(z{+}1,y)
\bar\Qoff^{\dot\alpha\alpha}
u_{\alpha}(h) }
\over
{ \bra{h} y{+}1,\ldots,n{-}1 \ket{n}
  \bra{n} 1,2,\ldots,z \ket{h} }
}
Y(\Qoff;z{+}1,\ldots,y) .
\label{PSIsubbed}
\eeq
Since there are no extra powers of $\Qoff$ in the denominator
of~(\ref{PSIsubbed}),  we should use the regulated form
for $Y$ given in Eq.~(\ref{regulatedY})
and perform a $d$-dimensional integration.
The integrals that occur are the same as the ones discussed
in Ref.~\cite{fourthpaper}, and so we proceed immediately to
the integrated expression for the color-ordered amplitude
\beqa
\amp(1^{+},\ldots,n^{+}) =
-{
{ (-\sqrt2)^n }
\over
{ 48\pi^2 }
}
\sum_{x=1}^{n-2} &&
\sum_{y=x+1}^{n-1}
\sum_{z=0}^{x-1}
u^{\alpha}(h)
\kappa_{\alpha\dot\alpha}(x{+}1,y)
\bar k^{\dot\alpha\beta}_x
u_{\beta}(h)
\LF && \times
{
{ u^{\gamma}(h)
  \kappa_{\gamma\dot\gamma}(z{+}1,y)
  [ \bar\kappa(z{+}1,x{-}1)-\bar\kappa(x{+}1,y) ]^{\dot\gamma\delta}
  u_{\delta}(h) }
\over
{ \bra{n}1,\ldots,z \ket{h}
  \bra{h}z{+}1,\ldots,y \ket{h}
  \bra{h}y{+}1,\ldots,n{-}1 \ket{n} }
}.
\label{GluAllPlusStart}
\eeqa
The expression in Eq.~(\ref{GluAllPlusStart}) must be gauge
invariant by itself:  the color factor for this process
contains no symmetries which would cause different terms
within the permutation sum to mix.

We begin our demonstration that~(\ref{GluAllPlusStart}) is
indeed independent of the gauge spinor by isolating a
portion of the numerator:
\beq
\num_2 \equiv
- u^{\gamma}(h)
  \kappa_{\gamma\dot\gamma}(z{+}1,y)
  [ \bar\kappa(z{+}1,x{-}1)-\bar\kappa(x{+}1,y) ]^{\dot\gamma\delta}
  u_{\delta}(h).
\eeq
By writing
$\kappa(z{+}1,y) = \kappa(z{+}1,x{-}1) + k_x + \kappa(x{+}1,y)$
we obtain
\beqa
\num_2 && =
- u^{\gamma}(h)
  \kappa_{\gamma\dot\gamma}(x{+}1,y)
  \bar\kappa^{\dot\gamma\delta}(z{+}1,x{-}1)
  u_{\delta}(h)
+ u^{\gamma}(h)
  \kappa_{\gamma\dot\gamma}(z{+}1,x{-}1)
  \bar\kappa^{\dot\gamma\delta}(x{+}1,y)
  u_{\delta}(h)
\LF  &&  \quad
- u^{\gamma}(h)
  k_{x\gamma\dot\gamma}
  \bar\kappa^{\dot\gamma\delta}(z{+}1,x{-}1)
  u_{\delta}(h)
+ u^{\gamma}(h)
  k_{x\gamma\dot\gamma}
  \bar\kappa^{\dot\gamma\delta}(x{+}1,y)
  u_{\delta}(h)
\LF &&
= - u^{\gamma}(h)
  \kappa_{\gamma\dot\gamma}(x,y)
  \bar\kappa^{\dot\gamma\delta}(z{+}1,x{-}1)
  u_{\delta}(h)
- u^{\gamma}(h)
  \kappa_{\gamma\dot\gamma}(x{+}1,y)
  \bar\kappa^{\dot\gamma\delta}(z{+}1,x)
  u_{\delta}(h).
\label{N2rewritten}
\eeqa
Since the two terms appearing in the last
line of~(\ref{N2rewritten}) differ
by only a shift in $x$ of 1 unit,
for the moment we need only discuss operations
on one of the two terms.

Considering then, the contribution from the first term, we
have
\beqa
\amp_1  \equiv
-{
{ (-\sqrt2)^n }
\over
{ 48\pi^2 }
}
\sum_{x=1}^{n-2} &&
\sum_{y=x+1}^{n-1}
\sum_{z=0}^{x-1}
u^{\alpha}(h)
\kappa_{\alpha\dot\alpha}(x{+}1,y)
\bar k^{\dot\alpha\beta}_x
u_{\beta}(h)
\LF && \times
{
{ u^{\gamma}(h)
  \kappa_{\gamma\dot\gamma}(x,y)
  \bar\kappa^{\dot\gamma\delta}(z{+}1,x{-}1)
  u_{\delta}(h) }
\over
{ \bra{n}1,\ldots,z \ket{h}
  \bra{h}z{+}1,\ldots,y \ket{h}
  \bra{h}y{+}1,\ldots,n{-}1 \ket{n} }
}.
\label{doZsum}
\eeqa
The sum on $z$ is very easy to perform
because of the identity
\beq
\sum_{i=a}^{b-1}
\link{i}{h}{i{+}1} = \link{a}{h}{b},
\label{linkID}
\eeq
a direct consequence of the Schouten identity.  Note that
we identify $k_0$ with $k_n$ in applying~(\ref{linkID})
to~(\ref{doZsum}), as implied by its denominator structure.
Interchanging the sum on $z$ with the implicit $\kappa$-sum
involving $z$ and using~(\ref{linkID}) yields
\beq
\amp_1  =
-{
{ (-\sqrt2)^n }
\over
{ 48\pi^2 }
}
\sum_{x=1}^{n-2}
\sum_{y=x+1}^{n-1}
u^{\alpha}(h)
\kappa_{\alpha\dot\alpha}(x{+}1,y)
\bar k^{\dot\alpha\beta}_x
u_{\beta}(h)
{
{ u^{\gamma}(h)
  \kappa_{\gamma\dot\gamma}(x,y)
  \bar\kappa^{\dot\gamma\delta}(1,x{-}1)
  u_{\delta}(k_n) }
\over
{ \bra{n}1,\ldots,y \ket{h}
  \bra{h}y{+}1,\ldots,n{-}1 \ket{n} }
}.
\label{doYsum}
\eeq

The sum on $y$ is a little more involved since there are
two implicit $\kappa$-sums involving $y$.  The relevant
structure to examine reads
\beqa
\sigma_y &\equiv&
\sum_{y=x+1}^{n-1}
\link{y}{h}{y{+}1}
[ u^{\alpha}(h)
  \kappa_{\alpha\dot\alpha}(x,y)
  \bar k^{\dot\alpha\beta}_x
  u_{\beta}(h) ]
[ u^{\gamma}(h)
  \kappa_{\gamma\dot\gamma}(x,y)
  \bar\kappa^{\dot\gamma\delta}(1,x{-}1)
  u_{\delta}(k_n) ]
\LF &=&
\sum_{y=x+1}^{n-1}
\sum_{a=x}^{n-1}
\sum_{b=x}^{y}
\link{y}{h}{y{+}1}
[ u^{\alpha}(h)
  k_{a\alpha\dot\alpha}
  \bar k^{\dot\alpha\beta}_x
  u_{\beta}(h) ]
[ u^{\gamma}(h)
  k_{b\gamma\dot\gamma}
  \bar\kappa^{\dot\gamma\delta}(1,x{-}1)
  u_{\delta}(k_n) ].
\eeqa
Interchanging the summations and using~(\ref{linkID}) to
do the sum on $y$ produces
\beqa
\sigma_y && =
{
{1}
\over
{\braket{h}{n}}
}
\sum_{a=x}^{n-1}
\sum_{b=x}^{a-1}
[ u^{\alpha}(k_n)
  k_{a\alpha\dot\alpha}
  \bar k^{\dot\alpha\beta}_x
  u_{\beta}(h) ]
[ u^{\gamma}(h)
  k_{b\gamma\dot\gamma}
  \bar\kappa^{\dot\gamma\delta}(1,x{-}1)
  u_{\delta}(k_n) ]
\LF && +
{
{1}
\over
{\braket{h}{n}}
}
\sum_{a=x}^{n-1}
\sum_{b=a}^{n-1}
[ u^{\alpha}(h)
  k_{a\alpha\dot\alpha}
  \bar k^{\dot\alpha\beta}_x
  u_{\beta}(h) ]
[ u^{\gamma}(k_n)
  k_{b\gamma\dot\gamma}
  \bar\kappa^{\dot\gamma\delta}(1,x{-}1)
  u_{\delta}(k_n) ].
\label{XtensionTime}
\eeqa
We extend the sum on $b$ appearing in the second term
of~(\ref{XtensionTime}) to the range $b \in [x,n{-}1]$.
This allows both sums to be performed in that piece.
The compensating term has precisely the same range as
the first term of~(\ref{XtensionTime}), and may be
combined with it by applying the Schouten identity.
The result of doing all of this is
\beqa
\sigma_y  = && -
\sum_{a=x}^{n-1}
\thinspace
  u^{\beta}(h)
  k_{x\beta\dot\alpha}
  \bar k^{\dot\alpha\gamma}_a
  \kappa_{\gamma\dot\gamma}(x,a{-}1)
  \bar\kappa^{\dot\gamma\delta}(1,x{-}1)
  u_{\delta}(k_n)
\LF && +
{
{1}
\over
{\braket{h}{n}}
}
[ u^{\alpha}(h)
  \kappa_{\alpha\dot\alpha}(x,n{-}1)
  \bar k^{\dot\alpha\beta}_x
  u_{\beta}(h) ]
[ u^{\gamma}(k_n)
  \kappa_{\gamma\dot\gamma}(x,n{-}1)
  \bar\kappa^{\dot\gamma\delta}(1,x{-}1)
  u_{\delta}(k_n) ].
\label{Ynearlydone}
\eeqa
The second term of~(\ref{Ynearlydone}) vanishes, since
\beqa
 u^{\gamma}(k_n)
  \kappa_{\gamma\dot\gamma}(x,n{-}1)
  \bar\kappa^{\dot\gamma\delta}(1,x{-}1)
  u_{\delta}(k_n)  & = &
- u^{\gamma}(k_n)
  \kappa_{\gamma\dot\gamma}(1,x{-}1)
  \bar\kappa^{\dot\gamma\delta}(1,x{-}1)
  u_{\delta}(k_n)
\LF  & = &
- \braket{n}{n} \kappa^2(1,x{-}1)
\LF & = & 0.
\eeqa
Inserting the non-vanishing contribution from~(\ref{Ynearlydone})
into~(\ref{doYsum})  and including the
piece generated from the second term of~(\ref{N2rewritten})
gives
\beqa
\amp(1^{+},\ldots,n^{+}) && =
{
{ (-\sqrt2)^n }
\over
{ 48\pi^2 }
}
\sum_{x=1}^{n-2}
\sum_{a=x+1}^{n-1}
{
{  u^{\beta}(h)
  k_{x\beta\dot\alpha}
  \bar k^{\dot\alpha\gamma}_a
  \kappa_{\gamma\dot\gamma}(x,a{-}1)
  \bar\kappa^{\dot\gamma\delta}(1,x{-}1)
  u_{\delta}(k_n) }
\over
{ \bra{n}1,\ldots,n \ket{h} }
}
\LF && +
{
{ (-\sqrt2)^n }
\over
{ 48\pi^2 }
}
\sum_{x=1}^{n-2}
\sum_{a=x+1}^{n-1}
{
{  u^{\beta}(h)
  k_{x\beta\dot\alpha}
  \bar k^{\dot\alpha\gamma}_a
  \kappa_{\gamma\dot\gamma}(x{+}1,a{-}1)
  \bar\kappa^{\dot\gamma\delta}(1,x)
  u_{\delta}(k_n) }
\over
{ \bra{n}1,\ldots,n \ket{h} }
}
\label{FinalStage}
\eeqa
for the color-ordered amplitude.

To execute the final stage of the reduction, it is sufficient
to consider only the numerators of~(\ref{FinalStage}).  Hence, we
define
\beqa
\num && \equiv
\sum_{x=1}^{n-2}
\sum_{a=x+1}^{n-1}
{  u^{\beta}(h)
  k_{x\beta\dot\alpha}
  \bar k^{\dot\alpha\gamma}_a
  \kappa_{\gamma\dot\gamma}(x,a{-}1)
  \bar\kappa^{\dot\gamma\delta}(1,x{-}1)
  u_{\delta}(k_n) }
\LF  && +
\sum_{x=1}^{n-2}
\sum_{a=x+1}^{n-1}
{  u^{\beta}(h)
  k_{x\beta\dot\alpha}
  \bar k^{\dot\alpha\gamma}_a
  \kappa_{\gamma\dot\gamma}(x{+}1,a{-}1)
  \bar\kappa^{\dot\gamma\delta}(1,x)
  u_{\delta}(k_n) }.
\label{LastNumerator}
\eeqa
Our first action is to write
$\kappa(x,a{-}1) = \kappa(1,a{-}1) - \kappa(1,x{-}1)$
to obtain
\beqa
\num && =
\sum_{x=1}^{n-2}
\sum_{a=x+1}^{n-1}
{  u^{\beta}(h)
  k_{x\beta\dot\alpha}
  \bar k^{\dot\alpha\gamma}_a
  \kappa_{\gamma\dot\gamma}(1,a{-}1)
  \bar\kappa^{\dot\gamma\delta}(1,x{-}1)
  u_{\delta}(k_n) }
\LF  && -
\sum_{x=1}^{n-2}
\sum_{a=x+1}^{n-1}
\thinspace
\kappa^2(1,x{-}1)
\thinspace
{  u^{\beta}(h)
  k_{x\beta\dot\alpha}
  \bar k^{\dot\alpha\gamma}_a
  u_{\gamma}(k_n) }
\LF  && +
\sum_{x=1}^{n-2}
\sum_{a=x+1}^{n-1}
{  u^{\beta}(h)
  k_{x\beta\dot\alpha}
  \bar k^{\dot\alpha\gamma}_a
  \kappa_{\gamma\dot\gamma}(x{+}1,a{-}1)
  \bar\kappa^{\dot\gamma\delta}(1,x)
  u_{\delta}(k_n) }.
\label{Ugh}
\eeqa
Using the Schouten identity on the first term of~(\ref{Ugh})
produces
\beqa
\num  = &&
-\sum_{x=1}^{n-2}
\sum_{a=x+1}^{n-1}
{  \braket{n}{h}  \thinspace
  \bar u_{\dot\alpha}(k_x)
  \bar k^{\dot\alpha\gamma}_a
  \kappa_{\gamma\dot\gamma}(1,a{-}1)
  \bar\kappa^{\dot\gamma\delta}(1,x{-}1)
  u_{\delta}(k_x) }
\LF  && +
\sum_{x=1}^{n-2}
\sum_{a=x+1}^{n-1}
{  u^{\beta}(k_n)
  k_{x\beta\dot\alpha}
  \bar k^{\dot\alpha\gamma}_a
  \kappa_{\gamma\dot\gamma}(1,a{-}1)
  \bar\kappa^{\dot\gamma\delta}(1,x{-}1)
  u_{\delta}(h) }
\LF  && -
\sum_{x=1}^{n-2}
\sum_{a=x+1}^{n-1}
\thinspace
\kappa^2(1,x{-}1)
\thinspace
{  u^{\beta}(h)
  k_{x\beta\dot\alpha}
  \bar k^{\dot\alpha\gamma}_a
  u_{\gamma}(k_n) }
\LF  && +
\sum_{x=1}^{n-2}
\sum_{a=x+1}^{n-1}
{  u^{\beta}(h)
  k_{x\beta\dot\alpha}
  \bar k^{\dot\alpha\gamma}_a
  \kappa_{\gamma\dot\gamma}(x{+}1,a{-}1)
  \bar\kappa^{\dot\gamma\delta}(1,x)
  u_{\delta}(k_n) }.
\label{FourTerms}
\eeqa
Since the first term of~(\ref{FourTerms}) produces a gauge-invariant
contribution to~(\ref{FinalStage}), we are led to suspect that
the last three terms sum to zero.  This is indeed the case, as
we shall now demonstrate.

Let $\zee$ equal the last three terms of~(\ref{FourTerms}).
We begin by extending the factor of $\kappa(x{+}1,a{-}1)$ appearing
in the last term of $\zee$
to $\kappa(1,a{-}1)$ and
compensating.  The result is
\beqa
\zee &=&
\sum_{x=1}^{n-2}
\sum_{a=x+1}^{n-1}
{  u^{\beta}(k_n)
  k_{x\beta\dot\alpha}
  \bar k^{\dot\alpha\gamma}_a
  \kappa_{\gamma\dot\gamma}(1,a{-}1)
  \bar\kappa^{\dot\gamma\delta}(1,x{-}1)
  u_{\delta}(h) }
\LF  &-&
\sum_{x=1}^{n-2}
\thinspace
\kappa^2(1,x{-}1)
\thinspace
{  u^{\beta}(h)
  k_{x\beta\dot\alpha}
  \bar\kappa^{\dot\alpha\gamma}(x{+}1,n)
  u_{\gamma}(k_n) }
\LF &+&
\sum_{x=1}^{n-2} \sum_{a=x+1}^{n-1} \thinspace
  u^{\beta}(h)
  k_{x\beta\dot\alpha}
  \bar k^{\dot\alpha\gamma}_a
  \kappa_{\gamma\dot\gamma}(1,a{-}1)
  \bar\kappa^{\dot\gamma\delta}(1,x)
  u_{\delta}(k_n)
\LF &-&
\sum_{x=1}^{n-2} \sum_{a=x+1}^{n-1} \thinspace
  \kappa^{2}(1,x) \thinspace
  u^{\beta}(h)
  k_{x\beta\dot\alpha}
  \bar k^{\dot\alpha\gamma}_a
  u_{\gamma}(k_n).
\label{Blah}
\eeqa
We now interchange the order of the sum on $x$ and the implicit
$\kappa$-sum involving $x$ that appears
in the third term of~(\ref{Blah}),
obtaining
\beqa
\zee = &-&
\sum_{x=1}^{n-2}
\thinspace
[\kappa^2(1,x{-}1)+\kappa^2(1,x)]
\thinspace
{  u^{\beta}(h)
  k_{x\beta\dot\alpha}
  \bar\kappa^{\dot\alpha\gamma}(x{+}1,n)
  u_{\gamma}(k_n) }
\LF &-&
\sum_{x=1}^{n-2}
\sum_{a=x+1}^{n-1}
{ u^{\delta}(h)
  \kappa_{\delta\dot\gamma}(1,x{-}1)
  \bar\kappa^{\dot\gamma\gamma}(1,a{-}1)
  k_{a\gamma\dot\alpha}
  \bar k^{x\dot\alpha\beta}
  u_{\beta}(k_n) }
\LF &+&
\sum_{x=1}^{n-2} \sum_{a=x+1}^{n-1} \thinspace
  u^{\beta}(h)
  \kappa_{\beta\dot\alpha}(x,a{-}1)
  \bar k^{\dot\alpha\gamma}_a
  \kappa_{\gamma\dot\gamma}(1,a{-}1)
  \bar k^{\dot\gamma\delta}_x
  u_{\delta}(k_n).
\label{BlahBlah}
\eeqa
In the second term of~(\ref{BlahBlah}) we extend
$\kappa(1,x{-}1)$ to $\kappa(1,a{-}1)$ and compensate:
\beqa
\zee = &-&
\sum_{x=1}^{n-2}
\thinspace
[\kappa^2(1,x{-}1)+\kappa^2(1,x)]
\thinspace
{  u^{\beta}(h)
  k_{x\beta\dot\alpha}
  \bar\kappa^{\dot\alpha\gamma}(x{+}1,n)
  u_{\gamma}(k_n) }
\LF &-&
\sum_{x=1}^{n-2}
\sum_{a=x+1}^{n-1}
{ \thinspace \kappa^2(1,a{-}1) \thinspace
  u^{\beta}(h)
  k_{a\beta\dot\alpha}
  \bar k^{\dot\alpha\gamma}_x
  u_{\gamma}(k_n) }
\LF &+&
\sum_{x=1}^{n-2}
\sum_{a=x+1}^{n-1}
{ u^{\delta}(h)
  \kappa_{\delta\dot\gamma}(x,a{-}1)
  \bar\kappa^{\dot\gamma\gamma}(1,a{-}1)
  k_{a\gamma\dot\alpha}
  \bar k^{\dot\alpha\beta}_x
  u_{\beta}(k_n) }
\LF &+&
\sum_{x=1}^{n-2} \sum_{a=x+1}^{n-1} \thinspace
  u^{\beta}(h)
  \kappa_{\beta\dot\alpha}(x,a{-}1)
  \bar k^{\dot\alpha\gamma}_a
  \kappa_{\gamma\dot\gamma}(1,a{-}1)
  \bar k^{\dot\gamma\delta}_x
  u_{\delta}(k_n).
\label{BlahBlahBlah}
\eeqa
When the sum over $x$ is performed in the second term
of~(\ref{BlahBlahBlah}), it is seen that it partially
cancels the first term.
The last two terms may be combined
by noting that
\beqa
  \bar\kappa^{\dot\gamma\gamma}(1,a{-}1)
  k_{a\gamma\dot\alpha}
+ \bar k^{\dot\gamma\gamma}_a
  \kappa_{\gamma\dot\alpha}(1,a{-}1)
&=& 2k_a\cdot\kappa(1,a{-}1) \thinspace
  \delta^{\dot\gamma}_{\dot\alpha}
\LF &=&
[\kappa^2(1,a) - \kappa^2(1,a{-}1)]
 \thinspace \delta^{\dot\gamma}_{\dot\alpha}.
\eeqa
Thus
\beqa
\zee = &-&
\sum_{x=1}^{n-2}
\thinspace
\kappa^2(1,x)
\thinspace
{  u^{\beta}(h)
  k_{x\beta\dot\alpha}
  \bar\kappa^{\dot\alpha\gamma}(x{+}1,n)
  u_{\gamma}(k_n) }
\LF &+&
\sum_{x=1}^{n-2}
\sum_{a=x+1}^{n-1}
\thinspace
[ \kappa^2(1,a) - \kappa^2(1,a{-}1)]
\thinspace
{ u^{\delta}(h)
  \kappa_{\delta\dot\gamma}(x,a{-}1)
  \bar k^{\dot\gamma\gamma}_x
  u_{\gamma}(k_n) }.
\label{BlahBlahBlahBlah}
\eeqa
The two terms in the double sum appearing in~(\ref{BlahBlahBlahBlah})
may be combined by shifting the sum over $a$ by 1 unit in
one of the two pieces, yielding
\beqa
\zee = &-&
\sum_{x=1}^{n-2}
\thinspace
\kappa^2(1,x)
\thinspace
{  u^{\beta}(h)
  k_{x\beta\dot\alpha}
  \bar\kappa^{\dot\alpha\gamma}(x{+}1,n)
  u_{\gamma}(k_n) }
\LF &-&
\sum_{x=1}^{n-2}
\sum_{a=x+1}^{n-1}
\thinspace
\kappa^2(1,a)
\thinspace
{ u^{\delta}(h)
  k_{a\delta\dot\gamma}
  \bar k^{\dot\gamma\gamma}_x
  u_{\gamma}(k_n) }.
\label{BlahBlahBlahBlahBlah}
\eeqa
When we perform the sum over $x$ in the second term
of~(\ref{BlahBlahBlahBlahBlah}), we see that it exactly
cancels the first term, giving $\zee=0$, as promised.

Thus, the entire amplitude is generated from the first
term of~(\ref{FourTerms}), with the result
\beqa
\amphat&&(1^{+},\ldots,n^{+}) =
{ {i}\over{48\pi^2} }
(-g\sqrt2)^n
\negthinspace \negthinspace \negthinspace
\permsum{1}{n{-}1}
\negthinspace\negthinspace
tr \thinspace \{ \Omega[1,n] \}  \thinspace
\sum_{x=2}^{n-2} \sum_{y=x+1}^{n-1} \negthinspace
{
{ \bar k_x^{\dot\alpha\alpha}
  \kappa_{\alpha\dot\beta}(1,x)
  \bar\kappa^{\dot\beta\beta}(1,y)
  k_{y\beta\dot\alpha} }
\over
{ \bra{1} 2,\ldots,n \ket{1} }
}.
\LF &&
\label{GLUE++++}
\eeqa
As a check of~(\ref{GLUE++++}), we note that when we set the
color factor equal to unity and (numerically) perform the
permutation sum, we recover the $n$-photon result reported
in Ref.~\cite{fourthpaper}.

Recently, Bern, Dixon, and Kosower~\cite{String} published a
conjecture for a particular color-ordered subamplitude
for the scattering of $n$ like-helicity gluons via
a {\it gluon}\ loop.  It is possible to connect the
result derived here for diagrams containing a quark loop
to their result by use of
supersymmetry identities~\cite{SUSYrelations}.
The bottom line is that the result~(\ref{GLUE++++}) should
agree with the conjecture of Ref.~\cite{String} up to a trivial
multiplicative factor.  We have
verified that
the two expressions do indeed agree.  Thus, our calculation
may be taken as the proof of this conjecture.

Since we know the expression for the gluon current appearing
in~(\ref{GenStart}) when the gluon labeled ``$n$'' carries
negative helicity (see Appendix), we may repeat the
calculation to obtain $\amphat(1^{+},\ldots,(n{-}1)^{+},n^{-})$.
After a somewhat lengthy calculation we obtain
\beq
\amphat(1^{+},\ldots,(n{-}1)^{+},n^{-}) =
-{ {i}\over{48\pi^2} }
(-g\sqrt2)^n
\negthinspace \negthinspace \negthinspace
\permsum{1}{n{-}1}
{
{ tr \thinspace \{ \Omega[1,n] \}  }
\over
{ \bra{1} 2,\ldots,n\ket{1} }
}
\thinspace\thinspace
\Xi(1^{+},\ldots,(n{-}1)^{+},n^{-})
\eeq
where
\beqa
\Xi&&(1^{+},\ldots,(n{-}1)^{+},n^{-}) =
\LF &&
\quad
\sum_{x=2}^{n-2}
{ u^{\gamma}(k_n)
  \kappa_{\gamma\dot\gamma}(x,n)
  \bar k^{\dot\gamma\delta}_x
  u_{\delta}(k_n) }
[k_n+\kappa(1,x)]^2
\LF && \qquad \times
\Biggl\{
\linkstar{n{-}1}{n}{1}
+ \sum_{s=2}^x
u^{\lambda}(k_n)
\POLE{\lambda}{\tau}(n,1,\ldots,s)
u_{\tau}(k_n)
\Biggr\}
\LF && +
\sum_{x=3}^{n-2}
{ u^{\gamma}(k_n)
  \kappa_{\gamma\dot\gamma}(x,n)
  \bar k^{\dot\gamma\delta}_x
  u_{\delta}(k_n) }
\LF &&  \quad\enspace
\times
\Biggl\{
{
{\bar u_{\dot\alpha}(k_1)}
\over
{\tekarb{n}{1}}
}
[\bar k_n {+} \bar\kappa(1,x{-}1)]^{\dot\alpha\beta}
u_{\beta}(k_n)
+ [k_n{+}\kappa(1,x{-}1)]^2
\sum_{s=2}^{x-1}
u^{\lambda}(k_n)
\POLE{\lambda}{\tau}(n,1,\ldots,s)
u_{\tau}(k_n)
\Biggr\}
\LF && -
\sum_{x=2}^{n-2}
{ u^{\gamma}(k_n)
  \kappa_{\gamma\dot\gamma}(x,n)
  \bar k^{\dot\gamma\delta}_x
  u_{\delta}(k_n) }
\thinspace\thinspace
\kappa^2(x,n)
\LF && \qquad
\times
\Biggl\{
\linkstar{1}{n}{n{-}1}
+ \sum_{r=x}^{n-2}
u^{\lambda}(k_n)
\POLE{\lambda}{\tau}(n,n{-}1,\ldots,r)
u_{\tau}(k_n)
\Biggr\}
\LF && -
\sum_{x=2}^{n-3}
{ u^{\gamma}(k_n)
  \kappa_{\gamma\dot\gamma}(x,n)
  \bar k^{\dot\gamma\delta}_x
  u_{\delta}(k_n) }
\LF && \quad\enspace
\times
\Biggl\{
{
{\bar u_{\dot\alpha}(k_{n-1})}
\over
{\tekarb{n}{n{-}1}}
}
\bar\kappa^{\dot\alpha\beta}(x{+}1,n)
u_{\beta}(k_n)
+ \kappa^2(x{+}1,n)
\sum_{r=x+1}^{n-2}
u^{\lambda}(k_n)
\POLE{\lambda}{\tau}(n,n{-}1,\ldots,r)
u_{\tau}(k_n)
\Biggr\}
\LF && -
\sum_{x=2}^{n-3} \sum_{r=x+2}^{n-1} \sum_{s=1}^{x-1}
u^{\gamma}(k_n)
\kappa_{\gamma\dot\gamma}(x,r{-}1)
\bar k^{\dot\gamma\delta}_x
u_{\delta}(k_n)
{
{1}
\over
{\braket{r}{s}}
}
u^{\lambda}(k_n)
{\teh_{\lambda}}^{\tau}(r,\ldots,n,1,\ldots,s)
u_{\tau}(k_n)
\LF && \quad\enspace
\times
u^{\beta}(k_r)
\biggl[
\kappa_{\beta\dot\alpha}(x,r{-}1)
\bar\kappa^{\dot\alpha\alpha}(s{+}1,x{-}1)
+
\kappa_{\beta\dot\alpha}(x{+}1,r{-}1)
\bar\kappa^{\dot\alpha\alpha}(s{+}1,x)
\biggr]
u_{\alpha}(k_s)
\LF && -
\sum_{x=2}^{n-3} \sum_{a=x+1}^{n-2}
\sum_{r=a+1}^{n-1} \sum_{s=1}^{a-1}
{
{1}
\over
{\braket{r}{s}}
}
u^{\lambda}(k_n)
{\teh_{\lambda}}^{\tau}(r,\ldots,n,1,\ldots,s)
u_{\tau}(k_n)
\LF && \qquad
\times
u^{\delta}(k_n)
k_{x\delta\dot\gamma}
\bar k_a^{\dot\gamma\gamma}
\biggl[
\kappa_{\gamma\dot\beta}(x,a)
\bar\kappa^{\dot\beta\alpha}(s{+}1,x{-}1)
+\kappa_{\gamma\dot\beta}(x{+}1,a)
\bar\kappa^{\dot\beta\alpha}(s{+}1,x)
\biggr]
u_{\alpha}(k_s)
\label{GLUE+++-}
\eeqa
and the function $\teh$ (read ``double pi'') is a particular
combination of $\pole$'s defined in the Appendix.
Not all of the terms of~(\ref{GLUE+++-}) contribute
when $n=4$ or 5.

Note that
$\pole(n,1,\ldots,n{-}2)$ appears in
the first term  of~(\ref{GLUE+++-}).
This function contains $[k_n+\kappa(1,n{-}2)]^{-2} = k_{n-1}^{-2},$
which is singular for an on-shell gluon.  However, there is a
factor of $k_{n-1}^2$ available in the prefactor to cancel this
singularity.  The difficulty with $\pole(n,n{-}1,\ldots,2)$
in the third term is resolved in the same manner.

Eq.~(\ref{GLUE+++-}) agrees with the Bern and Kosower results
for $n=4$ and $n=5$\cite{BK}.  This is the first calculation
including the cases $n\ge6$.  Additional checks are
possible, however, by comparing to the previously obtained
results containing photons.
Since the $\gamma g \rightarrow gg\ldots g$ scattering
amplitude may be generated from an all-gluon amplitude
simply by replacing a color matrix by the identity,
we should have the relation
\beq
\cycsum{1}{n-1}\amp(1,\ldots,n{-}1,n_{\gamma})
=
\cycsum{1}{n-1}\amp(1,\ldots,n{-}1,n)
\label{OneGluonToPhoton}
\eeq
connecting the two different color-ordered amplitudes,
independent of the helicities of the gauge bosons.
In the like-helicity case,~(\ref{OneGluonToPhoton}) is
obviously satisfied since the color-ordered amplitudes
themselves are identical
[{\it cf.}\ Eqs.~(\ref{PhotonGluon+}) and~(\ref{GLUE++++})].
In the case of a negative helicity gluon becoming a negative
helicity photon however [Eq.~(\ref{GLUE+++-})
versus Eq.~(\ref{PhotonGluon+})], the sum indicated
in~(\ref{OneGluonToPhoton}) must actually be performed to see the
equality.
We have verified
that the required
agreement is indeed present.
One further check is
provided by
replacing all of the gluons by photons.
Again, when the appropriate additional sums are performed, we
find agreement
with the results
reported in Ref.~\cite{fourthpaper}.


\section{Conclusions}

In this paper we have applied the double-off-shell quark
current containing $n$ like-helicity gluons to the problem
of one loop QCD corrections.  We have found that it is
a relatively simple matter to obtain
the corrections to those amplitudes that
vanish at tree-level.  This is not surprising, since such
amplitudes should have a relatively simple form,
with  no cuts in the complex plane.
We have obtained
compact expressions for
the helicity amplitudes for photon-gluon
scattering, electron-positron annihilation to gluons, and
gluon-gluon scattering via a quark loop in the case of
like-helicity gluons.   In addition, the gluon-gluon
amplitude with a single gluon of opposite helicity in
addition to an arbitrary number of like-helicity gluons
has been computed, albeit in a somewhat more complicated
form.

Of course, any realistic cross-section computation
involves a complete set of helicity amplitudes, since
the gluons are never observed directly in the final state.
In principle, it is a straightforward matter to obtain
these amplitudes from the recursion relations, although
a significant amount of computational labor is required.
A particularly important
question to investigate is the mechanism
for explicitly canceling the infrared divergences at
loop level against the appropriate tree diagrams.
This is an absolute necessity if one is to develop
numerical methods for evaluating complete cross-sections.

A second issue which should be examined is the validity
of using the massless limit for the quarks.
Although this limit seems reasonable for the
light flavors, the large mass of the $t$-quark
may translate into significant (if not dominant)
mass effects.
Thus, an efficient means of handling
massive fermions, preferably within a recursive
framework, should be sought.

In spite of these loose ends, much progress has been
made in the evaluation of one-loop QCD processes.
Amplitudes that are unapproachable from a direct attack
starting with only the Feynman rules for the theory
have been obtained, and in a far more compact form than
would be expected given the myriad of Feynman diagrams
involved.  There are additional lessons to be learned
from continued investigation utilizing the powerful
combination of the spinor representation, color factorization,
and recursion relations.


\acknowledgments

I would like to thank T.--M. Yan,
E. Laenen,
and W.T.  Giele for useful discussions during the course of this
work.  I would also like to thank T.--M. Yan and W.T. Giele
for reading the manuscript prior to release.

This work was started at Cornell University, where it
was supported in part by the National Science Foundation.



\appendix
\section*{Currents with one off shell particle}

For convenience,
we record here some of the solutions
to the Berends and Giele recursion relations
for currents with one off-shell particle~\cite{BG}.

For the $n$-gluon current, we have the color factorization
\beq
\Jmshat^x(1,\ldots,n) =
2g^{n-1} \permsum{1}{n} tr(\Omega[1,n]T^{x})
\Jms(1,\ldots,n),
\label{factorizeglue}
\eeq
valid for an arbitrary helicity configuration.
Note that there is no zero-gluon current.

We now turn to
specific
color-ordered currents relevant to the amplitudes computed
here. For the case of $n$ like-helicity gluons,
Berends and Giele find~\cite{BG}
\beq
\Jms_{\alpha\dot\alpha}(1^{+},\ldots,n^{+})
=
(-\sqrt2)^{n-1}
{
{  u_{\alpha}(h) u^{\beta}(h)\kappa_{\beta\dot\alpha}(1,n) }
\over
{ \bra{h} 1,\ldots,n \ket{h} }
},
\label{Jallplus}
\eeq
where we use the same gauge momentum $h$ for all of the gluons.

We also require  color-ordered gluon currents
containing a single gluon of opposite helicity.  In this
case the form of the expression depends on the location of
the opposite helicity gluon in the argument list of $\Jms$.
For a lone negative helicity gluon we write:
\beq
\Jms_{\alpha\dot\alpha}(n^{-})  =
{
{ u_{\alpha}(k_n) \bar u_{\dot\alpha}(h) }
\over
\tekarb{n}{h}
},
\eeq
that is, its gauge momentum is $h$.
All of the positive helicity gluons use
the momentum $k_n$ of the negative helicity gluon as their gauge
momentum.
If the negative helicity gluon is first in the list, we have
\beqa
\Jms_{\alpha\dot\alpha}(n^{-},1^{+},\ldots,z^{+})  = &&
(-\sqrt2)^{z}
{
{ u_{\alpha}(k_n) u^{\beta}(k_n)
  [k_n + \kappa(1,z)]_{\beta\dot\alpha} }
\over
{ \bra{n} 1,\ldots,z \ket{n} }
}
\negthinspace
\sum_{s=1}^z
u^{\gamma}(k_n)
\POLE{\gamma}{\delta}(n,1,\ldots,s)
u_{\delta}(k_n),
\LF &&
\label{MinusFirst}
\eeqa
while if the negative helicity gluon is last we find
\beqa
\Jms_{\alpha\dot\alpha}((y{+}1)^{+},\ldots,(n{-}1)^{+},n^{-})  = &&
-(-\sqrt2)^{n-y-1}
{
{ u_{\alpha}(k_n) u^{\beta}(k_n)
  \kappa_{\beta\dot\alpha}(y{+}1,n) }
\over
{ \bra{n} y{+}1,\ldots,n{-}1 \ket{n} }
}
\LF && \quad\times
\sum_{r=y+1}^{n-1}
u^{\gamma}(k_n)
\POLE{\gamma}{\delta}(n,n{-}1,\ldots,r)
u_{\delta}(k_n).
\eeqa
The function $\pole$ appearing here is defined exactly as
in~(\ref{poledef}) with one exception.  When precisely two  massless
momenta appear as arguments of $\pole$, this function is
singular.  For this situation, we define
\beq
u^{\alpha}(k_n)
{\pole_{\alpha}}^{\beta}(n,j)
u_{\beta}(k_n)
\equiv
\linkstar{h}{n}{j}, \qquad k_j^2 = k_n^2 = 0.
\label{specialPOLE}
\eeq
This special definition does not apply if $k_n^2 \ne 0$, as
$\pole$ is a well-defined quantity in that case.

Finally, we present the expression for the case of the negative
helicity gluon appearing somewhere in the middle of the argument
list:
\beqa
\Jms_{\alpha\dot\alpha}
((y{+}1)^{+},&&\ldots,(n{-}1)^{+},n^{-},1^{+},\ldots,z^{+})
=
\LF && -
(-\sqrt2)^{n-y+z-1}
{
{ u_{\alpha}(k_n) u^{\beta}(k_n)
  [\kappa(y{+}1,n) + \kappa(1,z)]_{\beta\dot\alpha} }
\over
{ \bra{n} y{+}1,\ldots,n{-}1\ket{n}
  \bra{n} 1,\ldots,z \ket{n} }
}
\LF && \qquad\quad\times
\sum_{r=y+1}^{n-1}
\sum_{s=1}^z
\invlink{r}{n}{s}
u^{\gamma}(k_n)
{\teh_{\gamma}}^{\delta}(r,\ldots,n{-}1,n,1,\ldots,s)
u_{\delta}(k_n)
\label{MinusMiddle}
\eeqa
where we have defined
\beqa
{\teh}(r,\ldots,n{-}1,n,1,\ldots,s)
\equiv &&
{\pole}(r,\ldots,n{-}1,n,1,\ldots,s)
\LF &&
-{\pole}(r{+}1,\ldots,n{-}1,n,1,\ldots,s)
\LF &&
-{\pole}(s,s{-}1,\ldots,1,n,n{-}1,\ldots,r)
\LF &&
+{\pole}(s{-}1,s{-}2,\ldots,1,n,n{-}1,\ldots,r).
\eeqa
The version of Eq.~(\ref{MinusFirst}) with $h=k_1$ was presented
by Berends and Giele in Ref.~\cite{BG}.  The remaining color-ordered
currents containing a single opposite-helicity gluon
closely
resemble the results obtained for the modified gluon
current in Ref.~\cite{thirdpaper}.  This is not surprising, since
one way to obtain this current is to begin with
a double-off-shell
gluon current with all like helicities.
One may then put one of the two off-shell gluons
on shell, assigning it
negative helicity.
Indeed, the inductive proof that~(\ref{MinusMiddle})
satisfies the Berends and Giele recursion relation is
virtually
identical to the proof given in Ref.~\cite{thirdpaper},
the differences lying entirely within
the terms covered by~(\ref{specialPOLE}).
This proof is easily adapted to handle these terms.

It should be noted that even though $\teh(n{-}1,1,n)$ is apparently
dependent upon the gauge momentum $h$,
it actually contains
\beqa
u^{\gamma}(k_n)\POLE{\gamma}{\delta}(n,1)u_{\delta}(k_n)
-u^{\gamma}(k_n)\POLE{\gamma}{\delta}(n,n{-}1)u_{\delta}(k_n) &=&
\linkstar{h}{n}{1} - \linkstar{h}{n}{n{-}1}
\LF &=&
\linkstar{n{-}1}{n}{1},
\eeqa
which is
independent of $h$.  Consequently, the expressions given
in~(\ref{GLUE+++-}) and~(\ref{MinusMiddle})
are also independent of $h$.


\begin{figure}[h]

\vskip0.2cm

\caption[]{The basic diagram for
$\gamma g \rightarrow gg\ldots g$.  The blob represents
the double-off-shell quark current, consisting of the sum
of all tree graphs containing exactly $n{-}1$ external gluons.
\label{photonfig}
\hskip7.55cm}

\vskip1.0cm

\caption[]{The basic diagram for $n$-gluon scattering via a quark
loop.  The gluons have been labeled as suggested by the
color structure of the second term of~(\ref{ColorFactor}).
\hskip7.95cm}\label{gluegluefig}

\end{figure}


\begin{references}
\bibitem{other}  Additional references to the
vast literature on this subject may be found within
Refs.~\cite{ubiquitous} and~\cite{review}.  \label{Other}
\end{references}
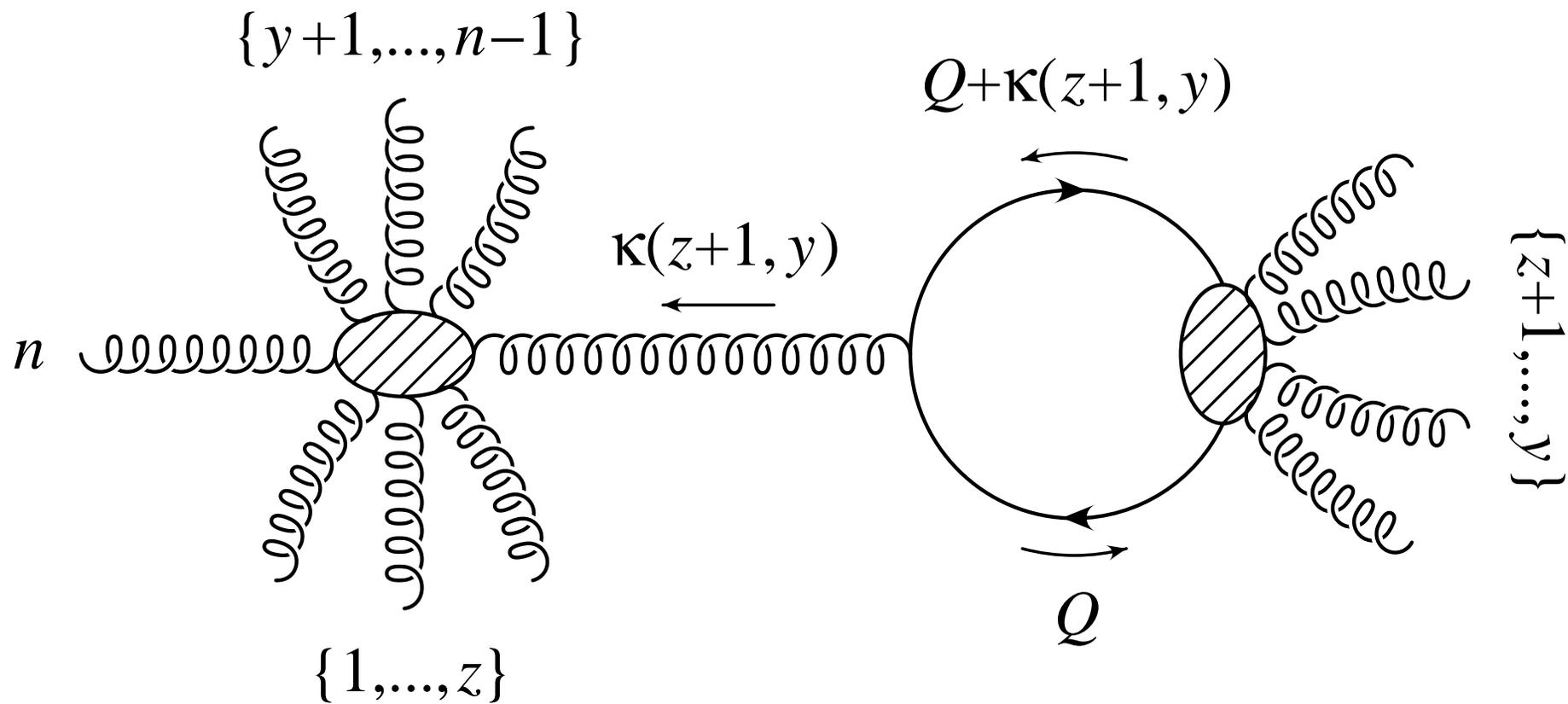
\end{document}